\documentclass[12pt]{iopart}

\newcounter{firstbib}

\usepackage{graphicx}
\usepackage{color}
\usepackage{upgreek}
\usepackage{iopams}
\expandafter\let\csname equation*\endcsname\relax
\expandafter\let\csname endequation*\endcsname\relax
\usepackage{amsmath}
\usepackage{array}
\usepackage{txfonts}
\usepackage{multirow}
\usepackage{fancyhdr}
\usepackage{multibib}
\newcommand{\ket}[1]{\left|#1\right\rangle}

\begin{document}

\title{Building Blocks of a Flip-Chip Integrated Superconducting Quantum Processor}
\author{
Sandoko Kosen\textsuperscript{1}\footnotemark,
Hang-Xi Li\textsuperscript{1}\footnotemark[\value{footnote}]\footnotetext{These authors contributed equally to this work},
Marcus Rommel\textsuperscript{1},
Daryoush Shiri\textsuperscript{1},
Christopher Warren\textsuperscript{1},
Leif Gr{\"o}nberg\textsuperscript{2},
Jaakko Salonen\textsuperscript{2},
Tahereh Abad\textsuperscript{1},
Janka Bizn{\'a}rov{\'a}\textsuperscript{1},
Marco Caputo\textsuperscript{2},
Liangyu Chen\textsuperscript{1},
Kestutis Grigoras\textsuperscript{2},
G{\"o}ran Johansson\textsuperscript{1},
Anton Frisk Kockum\textsuperscript{1},
Christian Kri\v{z}an\textsuperscript{1},
Daniel~P{\'e}rez Lozano\textsuperscript{1}\footnote{Present address: Imec, 3001 Leuven, Belgium},
Graham Norris\textsuperscript{3},
Amr Osman\textsuperscript{1},
Jorge~Fern{\'a}ndez-Pend{\'a}s\textsuperscript{1},
Alberto Ronzani\textsuperscript{2},
Anita Fadavi Roudsari\textsuperscript{1},
Slawomir Simbierowicz\textsuperscript{2}\footnote{Present address: Bluefors Oy, 00370 Helsinki, Finland},
Giovanna Tancredi\textsuperscript{1},
Andreas Wallraff\textsuperscript{3,4},
Christopher Eichler\textsuperscript{3},
Joonas Govenius\textsuperscript{2},
Jonas Bylander\textsuperscript{1}}

\address{\textsuperscript{1}Chalmers University of Technology, 412 96 Gothenburg, Sweden}
\address{\textsuperscript{2}VTT Technical Research Centre of Finland, FI-02044 VTT, Finland}
\address{\textsuperscript{3}ETH Z{\"u}rich, CH-8093 Z{\"u}rich, Switzerland}
\address{\textsuperscript{4}Quantum Center, ETH Z{\"u}rich, CH-8093 Z{\"u}rich, Switzerland}

\begin{abstract}
We have integrated single and coupled superconducting transmon qubits into flip-chip modules. Each module consists of two chips --- one quantum chip and one control chip --- that are bump-bonded together. We demonstrate time-averaged coherence times exceeding $90\,\upmu$s, single-qubit gate fidelities exceeding 99.9\%, and two-qubit gate fidelities above 98.6\%. 
We also present device design methods and discuss the sensitivity of device parameters to variation in interchip spacing.
Notably, the additional flip-chip fabrication steps do not degrade the qubit performance compared to our baseline state-of-the-art in single-chip, planar circuits.
This integration technique can be extended to the realisation of quantum processors accommodating hundreds of qubits in one module as it offers adequate input/output wiring access to all qubits and couplers.

\end{abstract}

%\maketitle

\section{Introduction}

The realisation of superconducting quantum processors with arrays of increasing numbers of qubits faces several interesting engineering and physics challenges \cite{Arute2019, Jurcevic2021, Gong2021}. From the hardware perspective, scaled-up circuit designs and fabrication processes must not degrade the device performance, which otherwise would adversely affect the fidelity of quantum algorithms. This already becomes non-trivial at the scale of dozens of interconnected qubits, which requires intricate routing of the input/output microwave circuitry.

In conventional devices fabricated on a single chip, one approach to routing signal lines is to implement signal crossovers using superconducting air-bridges \cite{Dunsworth2018,Andersen2021,Marques2021}. Furthermore, to minimise signal crosstalk, transmission lines can be enclosed in elongated `tunnels', which connect the ground planes on either side of the lines \cite{Dunsworth2018}.  
While feasible in small-scale circuits, these techniques alone seem insufficient for scaling up further. Monolithic integration, featuring buried multi-layer superconducting wiring, is an appealing solution \cite{Rosenberg2020}; however, this technology has not been demonstrated in coexistence with high-performance superconducting qubits. 

Two other scalable 3D-integration approaches are multi-chip (flip-chip or interposer chip) circuits and out-of-plane wiring \cite{Rosenberg2017,Gold2021,Arute2019,Jurcevic2021,Conner2021,Li2021a,Wu2021,Bronn2018,Bejanin2016,Tabuchi2019,Rahamim2017}. Both approaches have the advantage that the chip hosting the quantum circuitry can be fabricated separately, with minimal extra processing that risks degrading qubit performance.

Multi-chip implementations \cite{Rosenberg2017,Gold2021,Arute2019,Jurcevic2021,Conner2021,Li2021a,Wu2021} typically separate the quantum and the input/output wiring circuitry onto different chips. These chips are then connected to each other in a multi-chip module by flip-chip bump-bonding techniques. This enables flexible signal routing from the perimeter of the wiring chip to the capacitive, inductive, or galvanic point of contact with the quantum chip. Combined with superconducting through-silicon-via technology for grounding and signal routing \cite{Yost2020}, this technique would enable higher-density wiring with connections from the entire back plane of the wiring chip, rather than solely from the perimeter.

Out-of-plane wiring implementations eliminate the need for multiple chips, but instead deploy, e.g., spring-loaded \cite{Bronn2018,Bejanin2016,Tabuchi2019} or coaxial \cite{Rahamim2017} pins that approach the chip perpendicularly. Such implementations yield direct access to components within the two-dimensional qubit array, obviating the need of routing from the edges of the chip.

\begin{figure}[!ht]
    \centering
    \includegraphics[width=\linewidth]{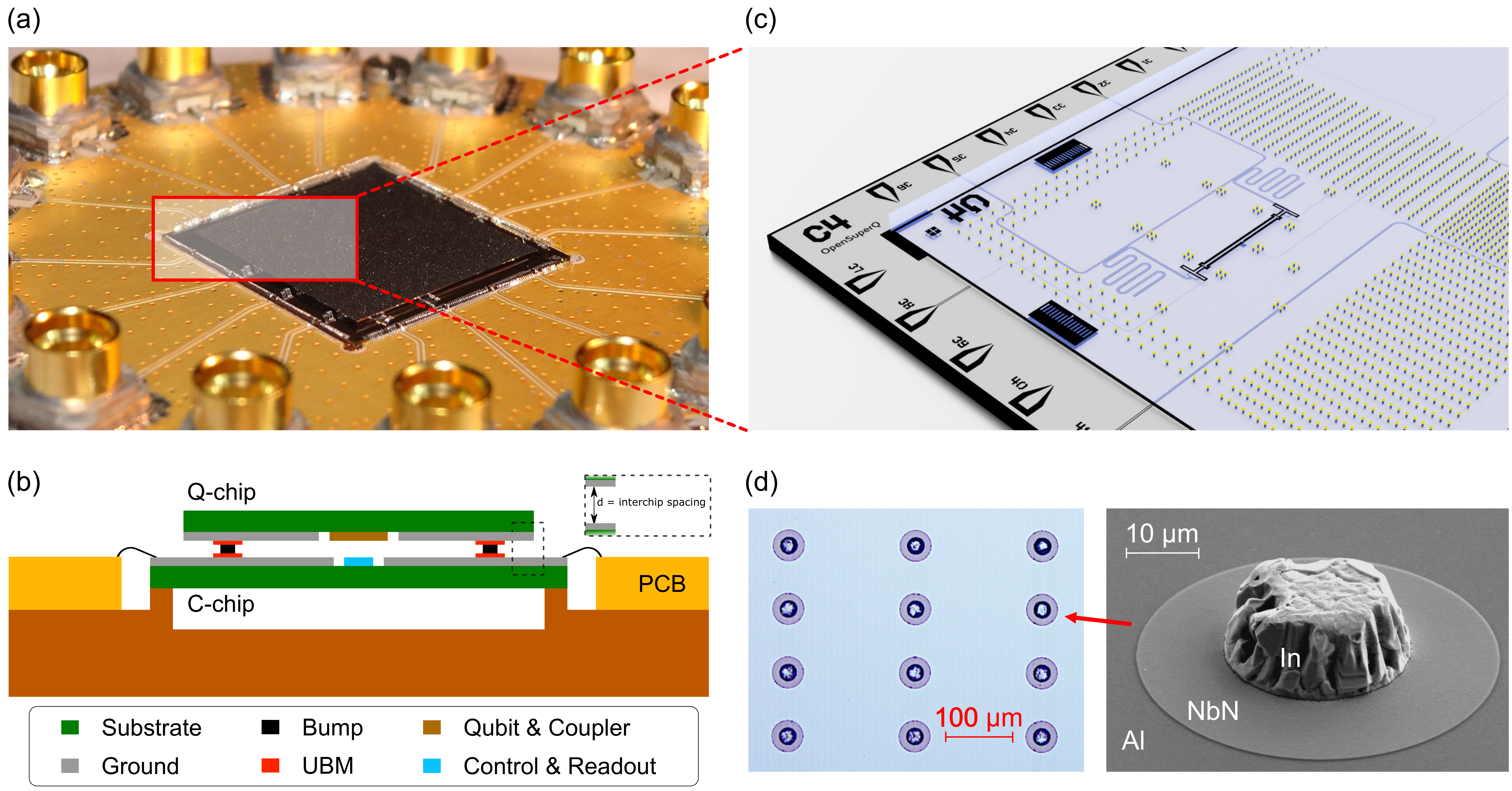}
    \caption{\textbf{Flip-chip module.} (a)\,Photograph of a flip-chip module within the centre cut-out of a printed circuit board (PCB), mounted in a microwave package (lid not shown). 
    (b)\,Simplified cross-sectional illustration of the flip-chip module (not to scale). The quantum chip (Q-chip) hosts qubits and couplers, i.e.,\@ any elements containing Josephson junctions. The control chip (C-chip) hosts the input/output wiring, i.e.,\@ control lines (XY and Z), readout resonators, and readout feedlines. The two chips are separated by arrays of bumps that provide galvanic connection and mechanical support. The module is mounted into a copper sample box with a recess below the chip, and the C-chip is wire-bonded to the PCB.
    (c)\,Illustrated 3D model of the flip-chip module with the substrate of the Q-chip rendered transparent. 
     (d)\,Images of the the bump layer (In), the under-bump metallisation layer (NbN), and the wiring layer (Al) prior to bonding. Bumps have the same layout on both chips. The image on the left (right) side was taken using an optical (a scanning electron) microscope.\label{fig1}}
\end{figure}

In this paper, we demonstrate superconducting quantum devices in a scalable architecture by integrating them into a flip-chip module comprised of two silicon chips, which we denote as the control (C) chip and the quantum (Q) chip, see fig.\,\ref{fig1}(a--c). The device consists of fixed-frequency aluminium transmon qubits and a flux-tunable parametric coupler on the Q-chip. These components are capacitively and inductively coupled to the control lines (XY and Z, respectively) and qubit-readout resonators located on the C-chip. The attained qubit performance is near the state of the art for flip-chip devices. We characterise single-qubit and two-qubit (CZ) gates with average fidelities of 99.97\,\% and 98.66\,\%, respectively. The measured average $T_1$ relaxation time in single-qubit devices is as high as $110\,\upmu$s, which, notably, is not degraded compared to our baseline single-chip devices \cite{Burnett2019,Bengtsson2020,Osman2021} and comparable to state-of-the-art 3D-integrated devices reported by other groups \cite{Gold2021,Arute2019,Jurcevic2021,Niedzielski2019,Patterson2019,Spring2021,Li2021a,Gong2021}. Moreover, we discuss device parameter sensitivities due to variations in the interchip spacing.

%%%%% FIGURE 1
%%%%%%

\section{Design and Simulation Workflow \label{main:designsimulation}}
The design process of the devices in this work typically begins by determining the target parameters at the Hamiltonian level (qubit and coupler $|0\rangle$-to-$|1\rangle$ transition frequencies $f_{01}$, anharmonicities $\alpha$, readout frequencies $f_\mathrm{r}$, coupling rates $g$, Purcell-decay limit $T_\mathrm{p}$, etc.).
These target parameters are chosen to be similar to those of our standard single-chip devices \cite{Bengtsson2020}. This comparison enables us to benchmark flip-chip device performances, and also validate the accuracy of the device simulation and fabrication processes. 

Next, we employ ANSYS---a finite element electromagnetic simulation software suite---to find the right geometry that corresponds to the target parameters \cite{AnsysEM}. Specifically, the ANSYS HFSS Eigenmode solver is used to predict resonator frequencies, while the ANSYS Maxwell Electrostatic solver is used to determine capacitance values between the various elements. In the simulation software, the model consists of two silicon (Si) chips that are facing each other, similar to the model shown in fig.\,\ref{fig1}(b). The chips are separated by the interchip spacing $d$. The metallic layer of each chip is represented by a planar sheet (zero thickness) located directly on the surface of each chip. The sheet is then given the appropriate boundary condition (for Eigenmode simulations) or excitation mode (for Electrostatic simulations). We typically aim for a convergence criterion below 0.5\% for both solvers (see Section 2 of the Supplementary Materials for more details).

In practice, variations in the interchip spacing $d$ introduced by the flip-chip bonding process leads to deviations from the target device parameters. To understand the extent to which these would affect the device parameters, we first build a simulation model with a target interchip spacing (here $d_\mathrm{target}=8\,\upmu$m). Next, two additional simulations are performed for $d=d_\mathrm{target} \pm 1\,\upmu$m, i.e.,\@ $7$ and $9\,\upmu$m. This leads to a range of device parameters that can be compared with the target parameters. In addition, this enables us to anticipate changes in device parameters --- caused by variations in $d$ --- that can lead to degraded device performance. For instance, the increase in the coupling strength between a qubit and its readout resonator when $d<d_\mathrm{target}$ may lead to a Purcell-limited qubit $T_1$. In such a case, we will target lower coupling strength at $d_\mathrm{target}=8\,\upmu$m so that the qubit is not Purcell-limited for a smaller $d$ (at least when $d>7\,\upmu$m). Section \ref{main:disc:param} will discuss in more details the sensitivity of various parameters to variations in $d$.

\section{Fabrication \label{main:fab}}

The device fabrication is based on our standard qubit process at Chalmers (150\,nm-thick aluminium [Al] on 280\,$\upmu$m-thick high-resistivity intrinsic Si) \cite{Burnett2019,Bengtsson2020}. The quantum (Q) and control (C) chips are connected together into a module by bump-bonding a pattern of compressible pillars of superconducting indium (In). This provides mechanical interchip separation and galvanic connection to their respective ground planes. Indium has been shown to be compatible with superconducting qubit fabrication processes \cite{Rosenberg2017,Foxen2018}. An under-bump-metallisation (UBM) layer of superconducting NbN separates the Al film and In pillars. The UBM is meant to act as a diffusion barrier that prevents the formation of an Al--In intermetallic state \cite{Gordon1999} and to protect the Al film from corrosion during bump fabrication.
The bumps and UBM are shown in fig.\,\ref{fig1}(d). 

At Chalmers, we fabricated two 2-inch wafers, one containing four C-chips ($14.3\,\mathrm{mm}\times 14.3\,$mm) and the other containing four Q-chips ($12\,\mathrm{mm}\times 12\,$mm). 
First, an Al film was deposited on both wafers by e-beam evaporation.
Second, the UBM pads were deposited in a process comprised of sputtering of a $50$\,nm-thick NbN film on patterned resists followed by liftoff. 
Thereafter, the wiring layer was etched out of the Al film. 
As a final step, the Josephson junctions (JJs) were fabricated on the Q-wafer \cite{Burnett2019}.

At VTT, In pillars on both wafers were formed by evaporation of a thick film ($8\,\upmu$m) on optically-patterned single-layer resist with a sidewall profile optimised for liftoff. After liftoff and dicing, individual chips were bonded into modules by compression at room temperature, creating superconducting electrical contacts between the chips without degrading the Al tunnel junctions and electrodes.

Again at Chalmers, the flip-chip modules were wire-bonded with Al wire to a printed circuit board (PCB) within an engineered, connectorised microwave package initially designed at ETH (see photograph in fig.\,\ref{fig1}[a] and cross-section illustration in fig.\,\ref{fig1}[b]).

\section{Device Characterisation}

\subsection{Interchip Spacing, Chip Tilt, and Transition Temperatures\label{main:devchar:dist_tilt}}
In this section, we report on the statistical analysis of the interchip spacing and chip tilt achieved in our flip-chip modules, and the superconducting transition temperatures of the materials comprising the flip-chip stack.

Deviations from the targeted interchip spacing $d_\mathrm{target}$ or non-zero tilt between the nominally parallel chip surfaces can occur due to slight variations in the bump-bonding process between different runs. We characterise these deviations non-destructively at VTT, using a scanning electron microscope, by measuring the distance $z_i$ between the two surfaces at each of the four corners of the flip-chip module. The interchip spacing $d$ at the centre of each module is inferred from the average of these four values, i.e., $d=(\sum_{i=1}^4 z_i)/4$. The chip tilt $\Delta d$ is defined as the largest difference between $z_i$ of any two corners, i.e., $\Delta d = \mbox{max}(|z_i-z_j|)$ for $i\ne j$. The chip tilt angle $\Delta \theta$ is defined as the largest tilt angle measured between any two corners. Measurements of 17 flip-chip modules yield the following sample means and standard deviations: $d=(7.8\pm0.8)\,\upmu$m, $\Delta d=(1.7\pm1.0)\,\upmu$m, and $\Delta \theta=(126\pm76)\,\upmu$rad. We refer the reader to Section 3.1 of the Supplementary Materials for more details.

Non-superconducting materials and interfaces in the flip-chip connections is a potential source of non-negligible Joule heating, especially when used for delivering inter-chip flux bias currents, which can be in the range of milliamperes. Using four-point probe measurements, we characterise the resistance of a daisy chain of 1200 bonded bump pairs, where each unit of the daisy chain consists of a planar interconnecting segment of Al and a pair of bonded In bumps with NbN UBM layers. In order to determine the superconducting transition temperatures $T_\mathrm{c}$ of the In bumps, we also characterise a separate daisy chain structure that omits Al and has NbN in the interconnecting planar segments. The $T_\mathrm{c}$ of NbN, In, and Al are observed near $12\,$K, $3.3\,$K, and $1.2\,$K respectively, indicating that all part of the stack are superconducting at millikelvin temperatures. Below approximately $1.2\,$K, we observe no resistance above the noise floor of the measurement system, which sets an upper bound of $\lesssim50\,$n$\Omega$ per daisy chain unit. This measurement-setup-limited upper bound is comparable to the values reported in other works \cite{Rosenberg2017, Foxen2018}. Refer to Section 3.2 of the Supplementary Materials for the plots of temperature-dependent resistance values for the different material stacks and more details about the measurement apparatus.

\subsection{Control Elements and Gate Fidelities \label{main:gatefidelities}}

%%%%% FIGURE 2
\begin{figure}
    \centering
    \includegraphics{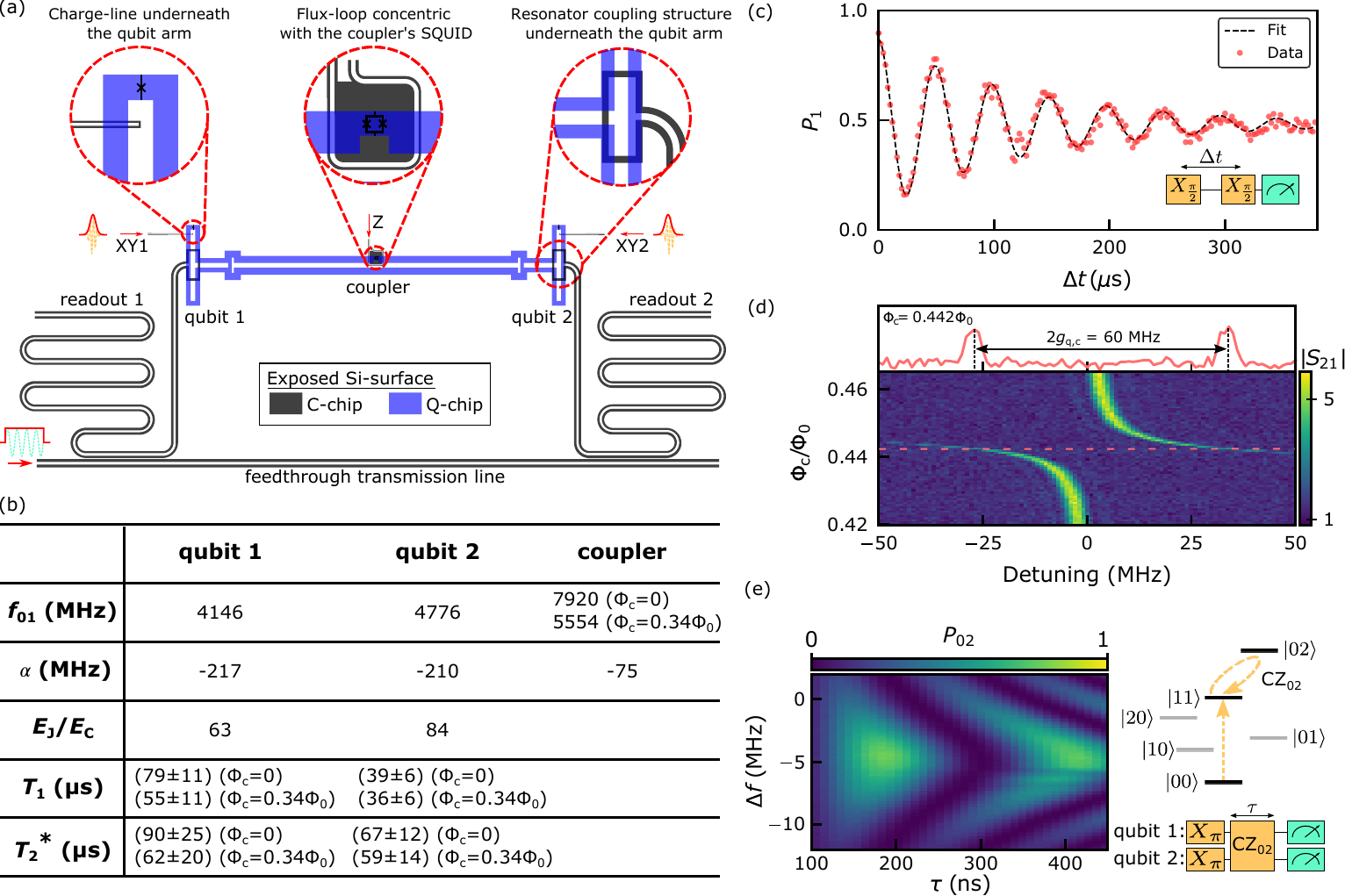}
    \caption{\textbf{Two-qubit flip-chip device.} (a)\,Illustration of two fixed-frequency transmon qubits and one frequency-tunable coupler, located on the Q-chip, and control lines (charge- or XY-line, flux- or Z-line), $\lambda/4$ readout resonators, and a feedthrough transmission line, located on the C-chip. The shaded area corresponds to the exposed silicon surface on each chip. The left inset shows the charge-lines (C-chip), opposite the qubit arm (Q-chip). The middle inset shows the flux-loop (C-chip), concentric with the superconducting quantum interference device (SQUID) loop of the coupler (Q-chip). The right inset shows the open-ended part of the readout resonator (C-chip), opposite the qubit (Q-chip). (b)\,Summary of device parameters (all were inferred from measurements except the anharmonicity $\alpha$ of the coupler). See Table 8 of the Supplementary Materials for more details. (c)\,Oscillation in the excited-state population of qubit~1, obtained by applying a Ramsey pulse sequence with 20\,kHz detuning, via XY1. (d)\,Avoided level crossing observed in the frequency spectroscopy of qubit~2 as the coupler is brought into resonance via the current applied to the Z-line. (e)\,Change in the population of state $|02\rangle$ vs modulation frequency detuning ($\Delta f$) and duration of the parametric modulation pulse ($\tau$). On the right, the energy diagram illustrates the CZ$_{02}$ transition and the gate sequence implemented in this experiment. The frequency detuning ($\Delta f$) is the difference between modulation frequency and resonant frequency of the $|11\rangle$-to-$|02\rangle$ transition, i.e., $f_\mathrm{CZ_{02}} = f_{01}(\mathrm{q2}) - f_{01}(\mathrm{q1}) + \alpha(\mathrm{q2})$. \label{fig2}}
\end{figure}
%%%%%%

In this section, we demonstrate the high-fidelity single-qubit and two-qubit gates driven by the control elements on the C-chip. Figure \ref{fig2}(a) illustrates a flip-chip device containing two fixed-frequency transmon qubits and a frequency-tunable coupler on the Q-chip. They face the control elements (XY1, XY2, Z), readout resonators (readout~1, readout~2), and a feedthrough transmission line on the C-chip. 
The XY-line (or charge-line) is an open-ended coplanar-waveguide transmission line, capacitively coupled to the qubit (see the leftmost inset of fig.\,\ref{fig2}[a]) and is used for driving qubit-state transitions. 
The Z-line (or flux-line) is a shorted loop, concentric with the coupler's superconducting quantum interference device (SQUID) axis (see the middle inset of fig.\,\ref{fig2}[a]), and is used to provide static and alternating magnetic flux to parametrically modulate the coupler and drive two-qubit gates. 
The readout elements are quarter-wavelength ($\lambda/4$) transmission-line resonators that are capacitively coupled to the qubits (see the rightmost inset of fig.\,\ref{fig2}[a]) and are also coupled to a feedthrough transmission line for multiplexed readout. 
The table in fig.\,\ref{fig2}(b) contains a basic summary of the device parameters. More details can be found in Section 4 of the Supplementary Materials.

We first demonstrate the functionality of the XY-line for coherent driving of the qubit. Figure \ref{fig2}(c) shows an oscillation in the first excited state population ($P_1$) of qubit~1 when a 20\,kHz detuned Ramsey pulse sequence is applied via XY1 \cite{Krantz2019}. 

Next, we turn our attention to the Z-line. By varying the direct current (DC) applied to the line, we change the magnetic flux $\Phi_\mathrm{c}$ threading the SQUID loop, which, in turn, changes the resonant frequency of the coupler ($f_\mathrm{c} = f_\mathrm{c0}\sqrt{|\cos{(\pi\Phi_\mathrm{c}/\Phi_0)}|}$, where $f_\mathrm{c0}$ is the zero-bias coupler frequency, $\Phi_0=h/2e$ is the flux quantum, $h$ is the Planck constant, and $e$ is the electron charge). When the coupler is tuned into resonance with the qubit, the two systems hybridise. This results in an avoided level crossing in the frequency spectroscopy data as shown in fig.\,\ref{fig2}(d) for qubit~2, yielding the coupler--qubit coupling strength for qubit~2, $g_\mathrm{q2,c}\sim 30\,$MHz. Similarly for qubit~1,  $g_\mathrm{q1,c}\sim27\,$MHz.

We focus on one of the two-qubit gates natively available in our coupled system---the controlled-Z (CZ) gate---which implements a $\pi$ phase shift on the joint qubit state $|\mathrm{q}_1\mathrm{q}_2\rangle=|11\rangle$ and leaves the other computational states unchanged. The CZ gate is implemented by parametric modulation of the coupler frequency via the Z-line \cite{Roth2017,McKay2016}. It brings the state on a round trip from $|11\rangle$ to $|02\rangle$ and back (referred to as the CZ$_{02}$ transition in the energy diagram of fig.\,\ref{fig2}(e)).

We bias the SQUID of the coupler at a non-zero flux offset (here, $\Phi_\mathrm{c}=0.34\Phi_0$), prepare both qubits simultaneously in the $|11\rangle$-state (via XY1, XY2), apply a pulsed, alternating-current (AC) modulation to the Z-line, and measure the joint probability of the $|02\rangle$-state (see the pulse sequence in fig.\,\ref{fig2}[e]). By varying both modulation frequency and CZ-pulse duration, we observe the expected oscillation in the population of the $|02\rangle$-state as shown in fig.\,\ref{fig2}(e).

%%%%% FIGURE 3
\begin{figure}
\includegraphics{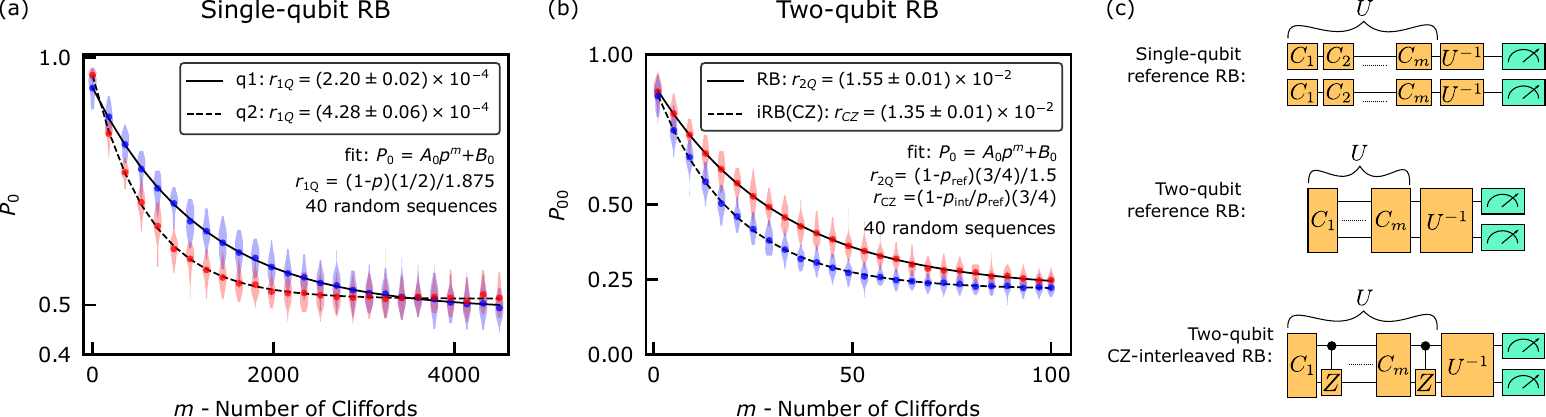}
\caption{\textbf{Characterisation of quantum gates} for the two-qubit flip-chip device shown in fig.\,\ref{fig2}. (a)\,Single-qubit reference randomised benchmarking (RB) performed simultaneously on both qubits. Single-qubit gates are 20\,ns wide pulses with a cosine envelope. The average physical gate errors on both qubits are below $r_\mathrm{1Q}\approx5\times10^{-4}$, equivalent to an average fidelity above $\mathcal{F}_\mathrm{1Q}\approx99.95\%$. The single-qubit interleaved RB results can be found in Table 9 of the Supplementary Materials. (b)\,Two-qubit reference and interleaved RB results for a 295\,ns-wide CZ-gate with a flat-top cosine envelope and virtual-Z gates (see main text for more details). The tunable coupler is biased at $\Phi=0.34\,\Phi_0$. The average error of the CZ-gate in this measurement run is $r_\mathrm{CZ}\approx 1.35\times10^{-2}$, equivalent to a CZ gate fidelity of $\mathcal{F}_\mathrm{CZ}\approx98.65\,\%$. (c)\,Gate sequences for the reference and interleaved RB experiments. \label{fig3}}
\end{figure}
%%%%%%

A common way to benchmark the performance of single-qubit and two-qubit gates is by performing randomised benchmarking (RB) experiments \cite{Magesan2011}. Reference RB experiments consist of the application of random Clifford sequences of varying lengths $m$ to the qubits which have been prepared in an initial state (typically the ground state), followed by an inverting gate to create an overall identity operation (see fig.\,\ref{fig3}[c]). Each Clifford is a combination of several physical gates from our gate set. The decay constant of the measured sequence fidelity versus $m$ provides an estimate of the average gate error \cite{Epstein2014,Magesan2012}.

Our single-qubit gate is implemented by applying a 20\,ns pulse with a cosine envelope, and is calibrated to maximise population transfer while minimising phase error \cite{Chen2018,Chow2010,Lucero2010}. Figure \ref{fig3}(a) shows results from single-qubit reference RB experiments performed simultaneously on both qubits. From this data, the average physical gate error of qubit~1 is $r_\mathrm{1Q}(\mathrm{q1})=(2.20\pm0.02)\times10^{-4}$, and of qubit~2 is $r_\mathrm{1Q}(\mathrm{q2})=(4.28\pm0.06)\times10^{-4}$, where the uncertainty represents the standard error of the fit. Furthermore, averaging the gate errors obtained from repeating single-qubit reference RB measurements over the course of $8\,$hours, without any additional gate recalibration in between, reveals similar results, $\overline{r}_\mathrm{1Q}(\mathrm{q1})=(2.3\pm0.1)\times10^{-4}$ for qubit~1 and $\overline{r}_\mathrm{1Q}(\mathrm{q2})=(4.2\pm0.1)\times10^{-4}$ for qubit~2, where the uncertainty is from the standard deviation of the spread (see fig.\,S4 in Section 4 of the Supplementary Materials). In terms of average physical gate fidelities, the 8-hour average values are $\overline{\mathcal{F}}_\mathrm{q1,meas}=1-\overline{r}_\mathrm{1Q}(\mathrm{q1})=(99.977\pm0.001)\%$ for qubit~1 and $\overline{\mathcal{F}}_\mathrm{q2,meas}=1-\overline{r}_\mathrm{1Q}(\mathrm{q2})=(99.958\pm0.001)\%$ for qubit~2.

The CZ gate consists of a 295\,ns flat-top pulse with a cosine-shaped rise and fall profile, complemented by virtual Z-rotations to both qubits. The latter serves to correct for additional dispersive shifts to both qubits introduced by the coupler during modulation \cite{Ganzhorn2020}. Figure \ref{fig3}(b) shows the result from implementing a reference two-qubit RB experiment reaching an average physical gate error of $r_\mathrm{2Q} = (1.55\pm0.01)\times10^{-2}$. As this reference RB sequence contains a mixture of single-qubit and CZ gates, a variant sequence is performed by interleaving each random Clifford with a CZ gate (see fig.\,\ref{fig3}[c] for the pulse sequence). The interleaved RB experiment can provide an estimate of the error per CZ gate, which in the case of fig.\,\ref{fig3}(b) is $r_\mathrm{CZ}=(1.35\pm0.01)\times10^{-2}$. Similarly, the average CZ error is characterised over the course of $10\,$hours. Without any gate-recalibration in-between, we obtain an average CZ error of $\overline{r}_\mathrm{CZ}=(1.34\pm0.08)\times10^{-2}$ or equivalently a CZ fidelity of $\overline{\mathcal{F}}_\mathrm{CZ,meas} = 1-\overline{r}_\mathrm{CZ}=(98.66\pm0.08)\%$ (see fig.\,S4 in Section 4 of the Supplementary Materials).

\subsection{Coherence Times \label{main:coherencetime}}
Another common benchmark is the qubit coherence times, which, in the absence of other errors, set a bound on the device performance.
In this section, we investigate whether the coherence time is substantially affected by the presence of another Si chip in close proximity and by the additional flip-chip fabrication process.

The table in fig.\,\ref{fig2}(b) summarises the coherence times of the two-qubit flip-chip device (shown in fig.\,\ref{fig2}[a]) obtained by interleaving measurements of relaxation times $T_1$ and Ramsey free-induction decay times $T_2^*$ simultaneously on both qubits, and repeatedly over a 36-hour period. 
In this device, qubits~1 and 2 exhibit average $T_1= 79\,\upmu$s and $39\,\upmu$s, respectively (at $\Phi_\mathrm{c}/\Phi_0=0$). Simple estimations of the limit imposed by decays via the readout resonator and the XY-line show that both qubits in this flip-chip module are indeed Purcell-decay limited, with qubit~2 being penalised more for having stronger coupling and smaller frequency detuning from its readout resonator (see discussion in Section 4.3 of the Supplementary Materials).

%%%%% FIGURE 4
\begin{figure}
    \centering
    \includegraphics{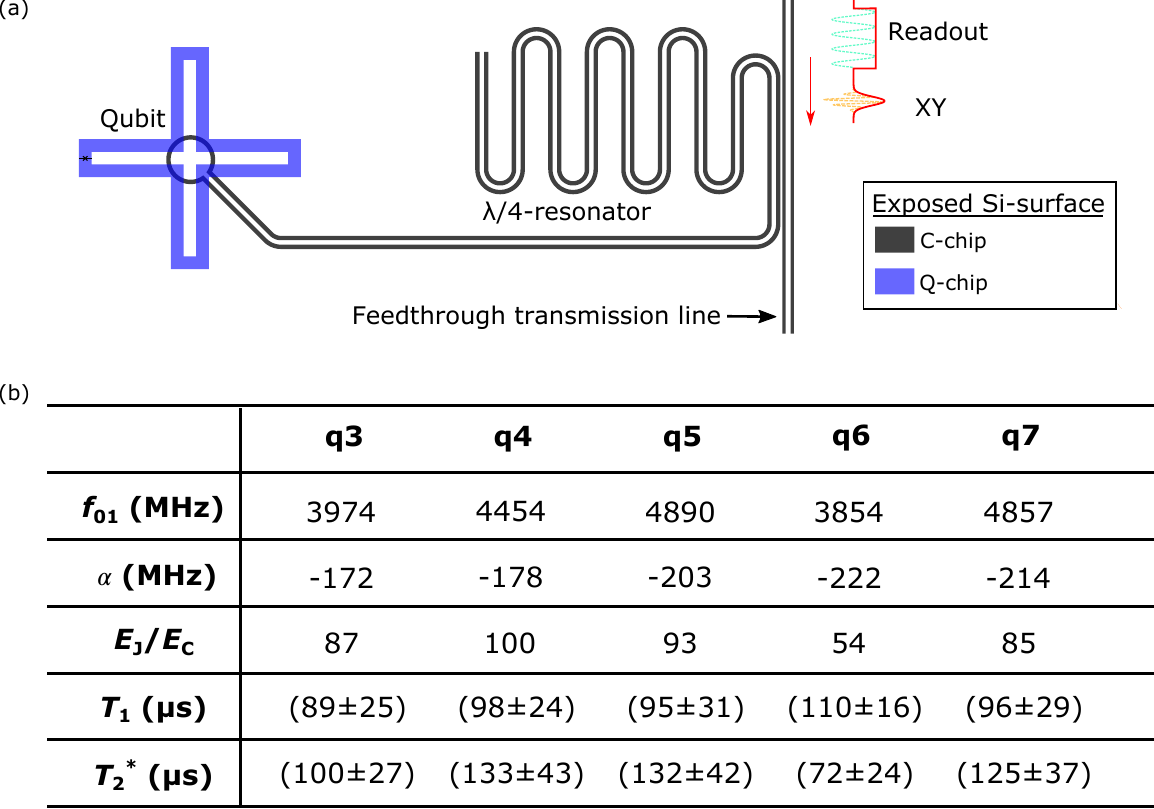}
    \caption{\textbf{Single-qubit coherence times in a flip-chip environment.} (a)\, Illustration of a fixed-frequency Xmon on the Q-chip, coupled capacitively to a $\lambda/4$-resonator on the C-chip. The drive (XY) and the readout pulses are coupled to the resonator via the feedthrough transmission line. (b)\,Summary of the measured parameters. The coherence times were obtained by interleaving measurements of $T_1$ and $T_2^*$ repeatedly over a 48-hour period. The table contains the mean and standard deviation of the data spread. Data were obtained from two separate flip-chip modules. Histogram and time series of $T_1$ and $T_2^*$ for each qubit can be found in fig.\,S6 of the Supplementary Materials. \label{fig4}}
\end{figure}
%%%%%%

To further investigate the level of qubit coherence that can be achieved in the flip-chip fabrication process outlined in Section \ref{main:fab}, we characterised the coherence times of single-qubit flip-chip devices in a configuration shown in fig.\,\ref{fig4}(a). In this simplified design, the `Xmon' \cite{Barends2013} qubit-state control and readout are both performed via the feedthrough transmission line that is coupled to the $\lambda/4$-resonator, i.e.,\@ without a separate XY-line. 
Basic parameters of the five single-qubit flip-chip Xmons characterised in this work are summarised in the table in fig.\,\ref{fig4}(b), with most exhibiting average values of $T_1$ and $T_2^*$ above $90\,\upmu$s. In contrast to the qubits in fig.\,\ref{fig2}(b), a simple estimation of the $T_1$-limit due to relaxation via the readout resonator shows that these qubits are not yet Purcell-decay limited by the coupling to the $\lambda/4$-resonator ($T_\mathrm{p} >160\,\upmu$s, see Section 5 of the Supplementary Materials).

\section{Discussion}
\subsection{Gate Fidelities and Coherence Times\label{main:disc:fidelities}}

Our demonstrated average gate fidelities in flip-chip are comparable to those obtained in single-chip devices \cite{Bengtsson2020} and those reported by other groups in flip-chip modules \cite{Jurcevic2021,Arute2019,Gold2021, Wu2021, Ye2021}. In particular, the short gate duration in Ref.\,\cite{Arute2019} enabled a CZ fidelity of almost 99.6\,\% in one qubit pair (fig.\,S17 in its Supplementary Materials), in a scheme that relies on having frequency-tunable qubits. The coherence-time limit to the gate fidelities in our case is estimated to be $\mathcal{F}_\mathrm{1Q, inc}\sim 99.98\%$ and $\mathcal{F}_\mathrm{CZ,inc}\sim 99.34\%$ according to Ref.\,\cite{Abad2021} (see Section~6 of the Supplementary Materials). This indicates that the measured average single-qubit gate fidelity ($\overline{\mathcal{F}}_\mathrm{1Q, meas}=(\overline{\mathcal{F}}_\mathrm{q1,meas}+\overline{\mathcal{F}}_\mathrm{q2,meas})/2\sim 99.97\%$) and the average CZ gate fidelity ($\overline{\mathcal{F}}_\mathrm{CZ,meas}= 98.66\%$) are still limited by both coherent and incoherent errors, with the single-qubit gate fidelities being closer to coherence-limited. Measures to improve the gate fidelities include decreasing the CZ gate duration, device redesign, and deploying more involved gate optimisation strategies \cite{Ganzhorn2020}.

Most of the single-qubit flip-chip devices in this work exhibit average $T_1>90\,\upmu$s with average $T_2^*>T_1$, which is not degraded compared to our previously reported single-chip qubit coherence times for similar device geometry \cite{Burnett2019,Osman2021}. Analysis of the participation ratio simulation results (Section 7 of the Supplementary Material) shows that the metallic layer of the second chip does cause a redistribution of electric energy from the substrate to the vacuum space in-between the chips while leaving that within the thin oxide interfaces largely unchanged. Therefore, we do not expect higher losses in flip-chip for the device geometry we use,  provided that the additional fabrication steps do not add significant lossy materials nor residues. This explains the similarity between the high coherence results in flip-chip and single-chip devices.
Our coherence times also compare favourably to values reported for flip-chip devices by other groups \cite{Zhang2020,Gold2021,Niedzielski2019,Jurcevic2021,Arute2019,Li2021a,Wu2021}.

Further work aiming to improve qubit coherence times in general will include device design and process development, in particular to reduce the loss contribution of dielectrics. 
Recently, transmons made of tantalum showed promisingly long $T_1$ as high as $\sim\!\!500\,\upmu$s in planar, single chips \cite{Place2021,Wang2021}. we note that Ref.\,\cite{Li2021a} reported on $T_1$ degradation of a ``floating" transmon design from $\sim\!\!160\,\upmu$s in the single-chip case down to $\sim\!60\,\upmu$s in the flip-chip case (see Table II \& III in Appendix A of Ref.\,\cite{Li2021a}), highlighting the difficulty of developing fabrication processes and designs compatible with preserved coherence.

In a scaled-up quantum processor consisting of a two-dimensional array of interconnected qubits, multiple signal lines would be routed close together on the C-chip and would also pass directly below the couplers on the Q-chip. Understanding the extent to which this would induce signal crosstalk and influence the device parameters by added capacitance, inductance, and dielectric loss participation is a matter of ongoing investigation.

\subsection{Parameter Sensitivity \label{main:disc:param}}
The ability to design and manufacture quantum processors according to the desired specification is of paramount importance in view of the resources required to fabricate and characterise them. One challenge with flip-chip integrated devices is that the interchip spacing $d$ plays an important role in determining the device parameters. The statistics taken for a small number of modules in Section \ref{main:devchar:dist_tilt} yield the following values for interchip spacing $d=(7.8\pm0.8)\,\upmu$m and chip tilt $\Delta d=(1.7\pm1.0)\,\upmu$m. This justifies the design strategy described in Section \ref{main:designsimulation} (design for $d_\mathrm{target}=8\,\upmu$m, and examine parameter changes at $d=d_\mathrm{target}\pm1\,\upmu$m), which partially allows us to include design margins toward deviations in device parameters due to module-to-module variations in $d$ and non-parallel chips (tilt, i.e., $\Delta d>0$) within each module.

To quantify the parameter sensitivity of the flip-chip device, we consider as an example the parameter variation in the two-qubit flip-chip device of fig.\,\ref{fig2}(a) due to a deviation of $d$ by $1\,\upmu$m from its target value. For larger deviations, see Section 4.3 of the Supplementary Material.

According to our electromagnetic simulations, a $1\,\upmu$m variation in $d$ results in a 2.6\% variation in the qubit self-capacitance.
This is in contrast to single-chip qubits, whose charging energy is determined entirely by lithography, with negligible variations. 
Such variations have implications for the achievable qubit-frequency precision; this is particularly important for multi-qubit processors with fixed-frequency qubits, for which the frequency allocation for neighbouring qubits has to be carefully designed to minimise crosstalk. 
Adding the variation due to Josephson-junction resistance variations in our current process \cite{Osman2021}, the estimated qubit-frequency variation for a 4\,GHz transmon qubit becomes 4.1\% (see Section 8 of the Supplementary Materials).

Other quantities are more strongly dependent on $d$, such as the coupling capacitance between XY-line and qubit or between readout resonator and qubit: our simulations indicate a 12\% change for a $1\,\upmu$m variation in $d$. This, in turn, affects the readout condition and the Purcell-decay limit imposed on the qubit. 
However, these couplings can be designed with a safe margin to anticipate variations induced during flip-chip bonding, as briefly described in Section \ref{main:designsimulation}. 

Likewise, to perform frequency-division multiplexed readout without crosstalk, resonators sharing the same feedthrough transmission line must be sufficiently separated in frequency. 
Module-to-module variations in $d$ result in off-target resonant frequencies: in the absence of chip tilt ($\Delta d=0$), the frequencies are shifted in the same direction, but in the presence of tilt they are shifted in opposite directions and may come too close. 
For the coplanar-waveguide resonators in fig.\,\ref{fig2}(a), our electromagnetic simulations indicate that a $1\,\upmu$m variation in $d$ results in a $2\%$ variation of the resonant frequency ($120\,$MHz for a 6\,GHz resonator). The frequencies must be allocated with this clearance in mind to ensure that frequency collisions between resonators due to chip tilt can be avoided.

Looking ahead, at least two possible improvements can be made to decrease the parameter sensitivity due to variations in $d$. The first is to implement hard-stop spacers to better control the resulting interchip spacing and tilt \cite{Niedzielski2019}. Another solution is to revise the device designs: for instance, a qubit facing a bare Si surface on the C-chip has a smaller relative contribution to its capacitance from the ground plane of the C-chip than one facing a superconducting ground plane, resulting in lower sensitivity to variations in $d$.

\subsection{Chip Deformation}
The chip can also be deformed during bonding. Such deformation would result in a non-linearly varying chip separation across the module, which, if not understood properly, could adversely affect our ability to accurately target specific device parameters in a large quantum processor. Unfortunately, such deformation cannot be reliably deduced solely from the measurement of chip separation on each corner of the module. A more thorough investigation would require a full imaging of the chip surface before and after bonding. A parallel investigation track would be to compare measured device parameters to simulation and infer the local separation between the two chips at the device location. 

\subsection{Choice of Target Interchip Spacing}
Two considerations guided our choice of target interchip spacing, $d_\mathrm{target}=8\,\upmu$m. Larger values of $d_\mathrm{target}$ result in device parameters that are less sensitive to deviations in $d$ (as shown in Section 4.3 of the Supplementary Material). On the other hand, the lithography and deposition of indium bumps becomes impractical for $d_\mathrm{target}>8\,\upmu$m. 
Therefore, we aimed for a target value of $8\,\upmu$m.

\section{Conclusion}
We have demonstrated the basic building blocks of a flip-chip integrated quantum processor and achieved transmon coherence times and quantum gate fidelities approaching the best flip-chip device performances reported in
literature \cite{Jurcevic2021,Arute2019,Wu2021}. 
A comparison of coherence times with those from in-house-fabricated single-chip devices indicates that the qubits are not degraded by the additional flip-chip fabrication steps at this level of performance. The pristine flip-chip environment demonstrated in this work is therefore ready to be used for the investigation and implementation of multi-qubit processors.  

\section{Acknowledgements}
We acknowledge the use of support and resources from Myfab Chalmers, and Chalmers Centre for Computational Science and Engineering (C3SE, partially funded by the Swedish Research Council through grant agreement no.\,2018-05973). This work was funded by the EU Flagship on Quantum Technology H2020-FETFLAG-2018-03 project 820363 OpenSuperQ and by the Knut and Alice Wallenberg (KAW) Foundation through the Wallenberg Centre for Quantum Technology (WACQT).

\section{Data Availability Statement}
The data that supports the findings of this study are available upon reasonable request from the authors.

\section*{References}
\providecommand{\newblock}{}

%\newpage

\title{ SUPPLEMENTARY MATERIALS for \newline ``Building Blocks of a Flip-Chip Integrated Superconducting Quantum Processor''}
\setcounter{section}{0}
\setcounter{figure}{0}
\renewcommand{\theequation}{S\arabic{equation}}
\renewcommand{\thefigure}{S\arabic{figure}}
\renewcommand{\thetable}{S\arabic{table}}
%%%%%%%%%% Prefix a "S" to all equations, figures, tables 
\section{Device Hamiltonian}
In this section, we establish the notation for parameters (closely following Ref.\,\cite{S_Krantz2019}) referred to in the rest of this supplementary material and the main text.

The Hamiltonian of a single transmon qubit $\mathcal{H}_\mathrm{q}$ is
\begin{equation}
\mathcal{H}_\mathrm{q} = 4E_\mathrm{C}\hat{n}^2 - E_\mathrm{J}\cos{\hat{\phi}},
\end{equation}
where $\hat{n}$ and $\hat{\phi}$ are the charge number and phase operators, $E_\mathrm{C}=e^2/2C_\mathrm{q}$ is the charging energy, $E_\mathrm{J}=hI_\mathrm{c}/(4\pi e)$ is the Josephson energy, and $I_\mathrm{c}$ is the junction critical current.

Expanding the $\cos{\hat{\phi}}$ term up to the second order $\hat{\phi}^2$, we obtain the Hamiltonian of a quantum harmonic oscillator $\mathcal{H}_\mathrm{o}= hf\hat{a}^\dagger\hat{a}$, where
\begin{align}
    \hat{\phi} = \left(\frac{2E_\mathrm{C}}{E_\mathrm{J}}\right)^{1/4}\left(\hat{a}^\dagger + \hat{a}\right),\quad \quad 
    \hat{n} = \frac{i}{2}\left(\frac{E_\mathrm{J}}{2E_\mathrm{C}}\right)^{1/4}\left(\hat{a}^\dagger - \hat{a}\right), \quad\quad [\hat{a},\hat{a}^\dagger]=1,
\end{align} 
and $f$ is the frequency. If we also retain the fourth-order diagonal term, which is sufficient for the purpose of establishing the notation in this text, then
\begin{equation}
    \frac{\mathcal{H}_\mathrm{q}}{h} \approx f_{01} \hat{a}^\dagger\hat{a} + \frac{\alpha}{2} \hat{a}^\dagger\hat{a}^\dagger\hat{a}\hat{a}\label{eq:suppl_singlequbitH},
\end{equation}
where $f_{01}$ is the transition frequency between the ground state $|0\rangle$ and the first excited state $|1\rangle$, and $\alpha=f_{12}-f_{01}$ is the qubit anharmonicity.

The Hamiltonian of a qubit coupled to its readout resonator (a harmonic oscillator with annihilation and creation operators $b$ and $b^\dagger$ such that $[\hat{b}, \hat{b}^\dagger]=1$) is
\begin{equation}
    \frac{\mathcal{H}_\mathrm{q,r}}{h} = \frac{\mathcal{H}_\mathrm{q}}{h} + f_\mathrm{r}  \hat{b}^\dagger\hat{b} + g_\mathrm{q,r}\left(\hat{a}^\dagger+\hat{a}\right)\left(\hat{b}^\dagger + \hat{b}\right),
\end{equation}
where $g_\mathrm{q,r}$ is the coupling strength between the qubit and the resonator. This Hamiltonian describes the coupled qubit-resonator system shown in fig.\,2 of the main text.

For a coupled two-qubit system such as the one shown in fig.\,2 of the main text, the Hamiltonian is
\begin{align}
    \frac{\mathcal{H}_\mathrm{q,r,c}}{h} =& \sum_{i=1,2}\frac{\mathcal{H}_\mathrm{q_i, r_i}}{h} + \frac{\mathcal{H}_\mathrm{c}}{h} + \sum_{i=1,2}g_\mathrm{q_i,c}\left(\hat{a}_\mathrm{q_i}^\dagger+\hat{a}_\mathrm{q_i}\right)\left(\hat{a}_\mathrm{c}^\dagger + \hat{a}_\mathrm{c}\right),
\end{align}
where $\mathcal{H}_\mathrm{c}$ is written in the form of eq.\,(\ref{eq:suppl_singlequbitH}) for the coupler, and $g_\mathrm{q_i,c}$ is the coupling strength between qubit $i$ and the coupler.

\section{Various Aspects of Device Parameter Simulation\label{SI:devsim}}
This section outlines several aspects of device simulation performed for the devices presented in this work. It complements the discussion presented in Section 2 of the main text.

\subsection{Resonator Frequency}

The resonance frequency of the resonator is simulated using the ANSYS HFSS Eigenmode solver. The substrate dielectric constant is set to 11.7. We typically aim for a convergence criterion in the form of \textit{Maximum Delta Frequency Per Pass} below 0.5\,\%.

\subsection{Resonator Coupling Quality Factor \label{SI:Qc}}
The coupling quality factor $Q_c$ of the resonator to the transmission line is simulated using the ANSYS HFSS Driven Modal solver with an initial mesh generated by the ANSYS HFSS Eigenmode solver \cite{S_AnsysEM}.

In the simulation model, the resonator is coupled to the transmission line. Each end of the transmission line is connected to the ground plane via a rectangular sheet. In the Eigenmode solver, the sheet is assigned the \textit{lumped RLC impedance} boundary condition; the resistance is set to 50\,$\Omega$, and the other two elements (capacitance and inductance) are set to \emph{None}. In the Driven Modal solver, the same rectangular sheet is instead assigned the \textit{Lumped Port} excitation (50\,$\Omega$ impedance). Both ground planes and transmission lines are set to have the \textit{perfect E} boundary condition.

The main simulation is run using the Driven Modal solver, where we use the \textit{Broadband Adaptive Solutions} option with a $\pm 100\,$MHz frequency range centred around the expected resonance frequency (\textit{Iterative Solver} with \textit{Mixed Order} basis functions). The frequency sweep type is \textit{Fast}, and we make sure that the frequency step is smaller than the expected resonance linewidth.

The Driven Modal solution setup is configured to use an initial mesh generated by the Eigenmode solver (\textit{Import Mesh} option). We require the Eigenmode solver to solve for at least two modes, one for the resonator and another one for the feedline. As only a rough initial mesh is required, we typically set the maximum number of passes in Eigenmode solver to be 10 (with a convergence criterion in the form of \textit{Maximum Delta Frequency Per Pass} of 1\,\%). Apart from this, we do not impose additional mesh refinement in the solver. For the Driven Modal solver, we typically aim for a convergence criterion in the form of \textit{Maximum Delta S} of 0.02.

It should be noted that it is possible to run the Driven Modal solver without importing an initial mesh from the Eigenmode solver. But we frequently find that the Driven Modal solver, when run alone, cannot catch the resonance of the resonator especially when the coupling between the resonator and the feedline becomes weaker.

After the Driven Modal simulation is completed, the $S_{21}$ parameter between the two ports of the feedline is fitted using the circle fit technique \cite{S_Probst2015}, which returns the internal quality factor $Q_\mathrm{i}$ (above $10^{7}$ as the dielectric loss tangent of the substrate was set to be $10^{-7}$), and the coupling quality factor $Q_\mathrm{c}$. The decay rate of the resonator to the transmission line, $\kappa_\mathrm{r}$, is calculated as $\kappa_\mathrm{r}=2\pi f_\mathrm{r}/Q_\mathrm{c}$.

The simulated resonance frequency $f_\mathrm{r}$ and the coupling quality factor $Q_\mathrm{c}$ are summarised in Table\,\ref{Table: Qc} (for the two-qubit flip-chip device) and Table\,\ref{Table: Qc_singlequbit} (for the single-qubit flip-chip devices). 

%%%%% TABLE %%%%%
\begin{table}[ht]
    \footnotesize
    \centering
    \begin{tabular}{|c | c | c|} 
    \hline
        Parameter & Res.\,1 & Res.\,2 \\ 
        \hline
        %[\heavyrulewidth]
        $Q_\mathrm{c,sim}$ & 12723 & 12086 \\ 
        \hline
        $f_\mathrm{r,sim}$\,(GHz) & 6.107 & 6.285 \\
        $\kappa_\mathrm{r,sim}/2\pi$\,(MHz) & 0.48 & 0.52  \\
        \hline
        $f_\mathrm{r,meas}$\,(GHz) & 6.210 & 6.370\\
        $\kappa_\mathrm{r,meas}/2\pi$\,(MHz) & 0.63 & 0.76 \\
        \hline
    \end{tabular}
\caption{\textbf{Two-qubit flip-chip device}---simulated and measured resonator $Q_\mathrm{c}$, $f_\mathrm{r}$, and $\kappa_\mathrm{r}$. 
The interchip spacing is set to 6.9\,$\upmu$m. $f_\mathrm{r,meas}$ and $\kappa_\mathrm{r,meas}$ are measured values from Table\,\ref{tableSI_1}. \label{Table: Qc}}
\end{table}
%%%%%%%%%%%%%%%

%%%%% TABLE %%%%%
\begin{table}[ht]
\footnotesize
    \centering
    \begin{tabular}{|c | c | c | c | c | c|} 
    \hline
        Parameter & Res.\,3 & Res.\,4 & Res.\,5 & Res.\,6 & Res.\,7  \\ 
        \hline
        %[\heavyrulewidth]
        $Q_\mathrm{c,sim}$ & 16347 & 13712 & 16415 & 15703 & 16471 \\ 
        \hline
        $f_\mathrm{r,sim}$\,(GHz) & 5.807 & 6.292 & 6.813 & 5.824 & 6.812\\
        $\kappa_\mathrm{r,sim}/2\pi$\,(MHz) & 0.36 & 0.38 & 0.42 & 0.37 & 0.41 \\
        \hline
        $f_\mathrm{r,meas}$\,(GHz) & 5.727 & 6.185 & 6.691 & 5.887 & 6.720\\
        $\kappa_\mathrm{r,meas}/2\pi$\,(MHz) & 0.23 & 0.19 & 0.49 & 0.25 & 0.37\\
        \hline
    \end{tabular}
\caption{\textbf{Single-qubit flip-chip device}---simulated and measured resonator $Q_\mathrm{c}$, $f_\mathrm{r}$, and $\kappa_\mathrm{r}$. 
The interchip spacing is set to 7.1\,$\upmu$m. $f_\mathrm{r,meas}$ and $\kappa_\mathrm{r,meas}$ are measured values from Table\,\ref{tableSI_3}. \label{Table: Qc_singlequbit}}
\end{table}
%%%%%%%%%%%%%%%

\subsection{Coupling Strength $g$\label{SI:g}} 
The coupling strengths of qubit--resonator $g_\mathrm{q,r}$ and qubit--coupler $g_\mathrm{q,c}$ are calculated using the capacitance values obtained from ANSYS Maxwell Electrostatics \cite{S_AnsysEM}.

Each metallic layer in the design environment is considered as one electrode of a capacitor. The capacitance matrix has diagonal terms that represent self capacitances and off-diagonal terms that represent coupling capacitance between different objects. The self capacitance of the transmon is used to determine its charging energy $E_\mathrm{c}$. For the devices presented in this work, we typically aim for an energy error convergence criterion below 0.5\,\%.
Tables \ref{tableSI:CapMxwl2} and \ref{tableSI:CapMxwl1} show the simulated capacitance matrix elements for the two-qubit and single-qubit flip-chip devices.

%%%%% TABLE %%%%%
\begin{table}[ht]
    \footnotesize
    \centering
    \begin{tabular}{|c | c | c | c|}
    \hline
         Parameter & q1 & q2 & coupler\\ 
        \hline
        %[\heavyrulewidth]
        $C_\mathrm{q,self}$ (fF) & 97.48 & 98.50 & 271.14\\ 
        $C^*_\mathrm{r,self}$ (fF) & 590.78 & 574.23 &  \\
        \hline
        %\midrule
        $C_\mathrm{q,r}$ (fF) & 8.37  & 8.37 &\\ 
        %\midrule
        $C_\mathrm{q,c}$ (fF) & 2.38 & 2.38 &\\
        %\midrule
        $C_\mathrm{q,XY}$ (fF) &  0.108 & 0.108 &\\ 
        \hline
    \end{tabular}
\caption{\textbf{Two-qubit flip-chip device}---simulated capacitance matrix elements. Interchip spacing is set to $6.9\,\upmu$m. \label{tableSI:CapMxwl2}}
\end{table}
%%%%%%%%%%%%%%%

%%%%% TABLE %%%%%
\begin{table}[ht]
    \footnotesize
    \centering
    \begin{tabular}{|c | c | c | c | c | c|} 
    \hline
         Parameter & q3 & q4 & q5 & q6 & q7 \\ 
        \hline
        %[\heavyrulewidth]
        $C_\mathrm{q,self}$(fF) & 127  & 114.99  & 103.09 & 99.53  & 97.54\\ 
        $C^*_\mathrm{r,self}$(fF) & 616.55  & 566.22  & 533.76 & 606.56  & 533.44\\ 
        \hline
        %\midrule
        $C_\mathrm{q,r}$(fF) & 10.3  & 8.29  & 6.89  & 7.76  & 6.45\\ 
        \hline
    \end{tabular}
\caption{\textbf{Single-qubit devices}---simulated capacitance matrix elements. The interchip spacing is set to 7.1\,um. \label{tableSI:CapMxwl1}}
\end{table}
%%%%%%%%%%%%%%%

The coupling capacitances of the qubit--coupler system $C_\mathrm{q,c}$ and qubit--resonator system $C_\mathrm{q,r}$ can be used to calculate the coupling coefficients $g_\mathrm{q,c}$ and $g_\mathrm{q,r}$, respectively. The coupling strength between two resonators is calculated with the aid of the following equation \cite{S_Sank2014},
\begin{equation}
g_{1,2} = \frac{\sqrt{f_{1}f_{2}}}{2}\frac{C_{1,2}}{\sqrt{C_\mathrm{1,self}C_\mathrm{2,self}}}\label{eq:g},
\end{equation}
where the indices 1 and 2 represent either the qubit, coupler, or resonator. $f_{1}$ and $f_{2}$ are the individual resonant frequency of the coupled systems, and $C_{1,2}$ is the coupling capacitance. Note that, for the resonator's capacitance, we use $C^*_\mathrm{r,self} = C_\mathrm{r,self}\cdot 2/\pi$, where $C_\mathrm{r,self}$ is its self-capacitance obtained from the Maxwell solver. The factor $ 2/\pi$ is to account for the fact that the simulation model assumes a constant voltage throughout the structure, while the actual resonator is a $\lambda/4$ resonator with a non-homogeneous field distribution.

Table\,\ref{tableSI:g_twoqubit} contains the calculated coupling strengths between qubit and resonator and coupler in the two-qubit flip-chip device.
Table\,\ref{tableSI:g_singlequbit} contains those for the single-qubit devices.

%%%%% TABLE %%%%%
\begin{table}[ht]
% \begin{ruledtabular}
    \footnotesize
    \centering
    \begin{tabular}{|c | c | c|}
    \hline
         Parameter & q1 & q2 \\ 
        \hline
        %[\heavyrulewidth]
        $g_\mathrm{q,r,sim}$ (MHz) & 90  & 97\\
        %\midrule
        $g_\mathrm{q,r,meas}$ (MHz) & 95  & 115\\
        \hline
        %\midrule
        $g_\mathrm{q,c,sim}$ (MHz) & 31  & 34\\
        %\midrule
        $g_\mathrm{q,c,meas}$ (MHz) & 27  & 30\\
        \hline
    \end{tabular}
% \end{ruledtabular}
\caption{\textbf{Two-qubit device}---simulated values of coupling strengths. The interchip spacing is set to 6.9\,um. $g_\mathrm{q,r,meas}$ and $g_\mathrm{q,c,meas}$ are measured values from Table\,\ref{tableSI_1}. \label{tableSI:g_twoqubit}}
\end{table}
%%%%%%%%%%%%%%%

%%%%% TABLE %%%%%
\begin{table}[ht]
\footnotesize
    \centering
    \begin{tabular}{|c|c|c|c|c|c|}
    \hline
         Parameter & q3 & q4 & q5 & q6 & q7 \\ 
        \hline
        %[\heavyrulewidth]
        $g_\mathrm{q,r,sim}$ (MHz) & 86  & 83  & 82  & 74 & 79\\ 
        $g_\mathrm{q,r,meas}$ (MHz) & 85  & 84 & 79 & 80  & 81\\ 
        \hline
    \end{tabular}
\caption{\textbf{Single-qubit devices}---simulated values of coupling strengths. The interchip spacing is set to 7.1\,um. $g_\mathrm{q,r,meas}$ are measured values from Table\,\ref{tableSI_3}. \label{tableSI:g_singlequbit}}
\end{table}
%%%%%%%%%%%%%%%

The coupling capacitance between the qubit and XY-line is also simulated by the Electrostatic solver, from which the quality factor of the coupling between the qubit and the XY-line $Q_\mathrm{c}(\mathrm{XY})$ can be extracted,
\begin{equation}
Q_\mathrm{c}(\mathrm{XY})=\frac{C_\mathrm{q,self}}{\omega_{01}C_\mathrm{q,XY}^2R_\mathrm{e}},\label{eq:Qext-XY}
\end{equation}
where $\omega_{01}=2\pi f_{01}$ is the angular frequency of the qubit and $R_\mathrm{e}$ is the resistance of external load. In our case, $R_\mathrm{e}=50\,\Omega$ is assumed to be the characteristic impedance of XY-line.

\subsection{Purcell-decay Limited Lifetime $T_\mathrm{p}$\label{SI:purcell}}

The Purcell lifetime due to decay via the readout resonator, i.e., $T_\mathrm{p}(\mbox{read})$, is given by Ref.\,\cite{S_Koch2007} as
\begin{equation}
    \frac{1}{T_\mathrm{p}\mbox{(read)}} = \kappa_\mathrm{r} \left( \frac{g_\mathrm{q,r}}{\Delta_\mathrm{q,r}} \right)^2 \label{eq:purcellreadout}, 
\end{equation}
where $\kappa_\mathrm{r}$ is the decay rate of the resonator to the transmission line, $g_\mathrm{q,r}$ is the coupling rate between the qubit and the resonator, and $\Delta_\mathrm{q,r}=f_{01}-f_\mathrm{r}$ is the qubit--readout detuning.

The Purcell lifetime due to decay via the XY-line, i.e., $T_\mathrm{p}(\mbox{XY})$, is given by Ref.\,\cite{S_Sank2014} as
\begin{equation}
\frac{1}{T_\mathrm{p}(\mathrm{XY})} = \frac{\omega_{01}}{Q_\mathrm{c}(\mathrm{XY})}= \frac{R_\mathrm{e} (\omega_{01}C_\mathrm{q,XY})^2}{C_\mathrm{q,self}},\label{eq:purcelldrive}
\end{equation}
where $R_\mathrm{e}$ is the resistance of the external load (assumed to be 50\,$\Omega$), $\omega_{01}$ is the angular frequency of the qubit $|0\rangle$-to-$|1\rangle$ transition, $C_\mathrm{q,XY}$ is the capacitive coupling between the XY-line and the qubit, and $C_\mathrm{q,self} =e^2/2E_\mathrm{c}$ is the qubit self-capacitance.

We take the combined Purcell decay rate as the sum of both rates: $1/T_\mathrm{p}(\mathrm{read+XY}) = 1/T_\mathrm{p}(\mathrm{read}) + 1/T_\mathrm{p}(\mathrm{XY})$.

\section{Measurements of Interchip Spacing, Chip Tilt, and Transition Temperatures}
\subsection{Interchip Spacing and Chip Tilt \label{SI:chiptiltrawdata}}
\begin{figure}[h!]
    \centering
    \includegraphics{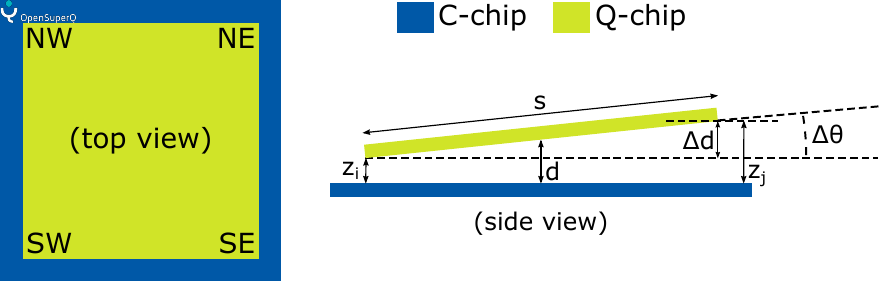}
    \caption{\textbf{Simplified illustration of the flip-chip module.}\label{figSI_chip}}
\end{figure}

Figure \ref{figSI_chip} shows the four corners of a flip-chip module labelled as: north east (NE), north west (NW), south east (SE), and south west (SW). The characterisation of each flip-chip module was performed by measuring the distance $z_i$ between the two surfaces of the chips at each of these four corners. The distance measurements were performed non-destructively using a scanning electron microscope.

%%%%% FIGURE 
%%%%%%

The interchip spacing $d$, defined as the distance between the surfaces of the two chips at the centre of each module, is inferred from mean distance across the four corners. The chip tilt $\Delta d$ is defined as the largest difference of $z_i$ between any two corners, i.e., $\Delta d = \mbox{max}(|z_i-z_j|)$ for $i\ne j$. This includes the tilt along the edges (NW-NE, NE-SE, SE-SW, SW-NW) and the diagonals (NW-SE, NE-SW). Furthermore, a tilt angle can be defined for each pair of corners, i.e., $\theta_{ij}=|z_i-z_j|/s$, where $s=12\,$mm ($16.97\,$mm) if the considered tilt is along the edge (the diagonal) of the module. The chip tilt angle $\Delta \theta$ is then defined as the largest tilt angle measured between any two corners, i.e., $\Delta \theta = \mbox{max}_{ij}\theta_{ij}$.

Table \ref{tableSI_chip} contains a summary of the measured distances $z_i$ of the four corners of each flip-chip module. There were 8 fabrication runs, producing 23 flip-chip modules in total. We only take into account measurements that were obtained using a scanning electron microscope (SEM). Therefore, we exclude the first generation FC1 as the distances were characterised using an optical microscope. We also exclude the fifth generation FC4a as the modules were not bonded with the proper tool. As a result, only 17 flip-chip modules are included in our statistical analysis of the interchip spacing and chip tilt.

From these data, we infer the following mean and standard deviation values: $d=(7.8\pm0.8)\,\upmu$m, $\Delta d=(1.7\pm1.0)\,\upmu$m, and $\Delta \theta=(126\pm76)\,\upmu$rad. 

%%%%% TABLE
\begin{table}[!h]
    \footnotesize
    \centering
    \begin{tabular}{ | c| c| c| c| c| c| c| c| c| c|}
    \hline 
    \multirow{2}{*}{\#} &
    \multirow{2}{*}{Module} &
    \multicolumn{4}{c|}{$z$ ($\upmu$m)} &
    $d$ &
    $\Delta d$ &
    $\Delta \theta$ &
    \multirow{2}{*}{Remarks} \\
    \cline{3-6} &
    & SE & NE & SW & NW & ($\upmu$m) & ($\upmu$m) & ($\upmu$rad) & \\
    \hline
    FC1 & CQ1 & 8.5 & 7.8 & 8 & 8.5 & 8.2 & 0.7 & 58 & 
    \multirow{4}{*}{\parbox{2cm}{optical \\ microscope}}\\
     & CQ2 & 8.3 & 8.0 & 8.0 & 7.0 & 7.8 & 1.3 & 83 & \\
      \cline{3-9}
     & C3,Q3& \multicolumn{7}{c|}{not bonded} & \\
      \cline{3-9}
     & CQ4 & 8.0 & 6.3 & 7.5 & 7.0 & 7.2 & 1.7 & 141 & \\ \hline
    
    FC2 & CQ1 & 7.77 & 7.81 & 8.44 & 8.48 & 8.13 & 0.71 & 56 & 
    \multirow{4}{*}{\parbox{2cm}{SEM}} \\
     & CQ2 & 8.05 & 8.28 & 7.89 & 8.12 & 8.09 & 0.39 & 23 & \\
      \cline{3-9}
     & C3,Q3& \multicolumn{7}{c|}{not bonded} &\\
      \cline{3-9}
     & CQ4 & 8.55 & 8.75 & 7.81 & 8.01 & 8.28 & 0.94 & 62 & \\ \hline
    
    FC3a & CQ1 & 7.43 & 7.66 & 8.83 & 9.06 & 8.25 & 1.63 & 117 & 
    \multirow{4}{*}{\parbox{2cm}{SEM}} \\
     & CQ2 & 11.33 & 9.22 & 9.14 & 7.03 & 9.18 & 4.3 & 253 & \\
      \cline{3-9}
     & C3,Q3& \multicolumn{7}{c|}{not bonded} & \\
      \cline{3-9}
     & CQ4 & 8.88 & 9.14 & 9.31 & 9.57 & 9.23 & 0.69 & 41 & \\ \hline 
     
    FC3b & CQ1 & 8.62 & 8.35 & 7.65 & 7.38 & 8.00 & 1.24 & 81 & 
    \multirow{4}{*}{\parbox{2cm}{SEM}} \\
     & CQ2 & 7.29 & 8.36 & 7.61 & 8.58 & 7.94 & 1.29 & 89 & \\
      \cline{3-9}
     & C3,Q3& \multicolumn{7}{c|}{not bonded} &\\
      \cline{3-9}
     & CQ4 & 7.22 & 6.88 & 8.05 & 7.71 & 7.47 & 1.17 & 69 & \\ \hline 
    
    FC4a & CQ1 & 9.29 & 9.49 & 14.30 & 10.90 & 11.00 & 5.01 & 418 & 
    \multirow{4}{*}{\parbox{2cm}{SEM, \\improper tool}} \\
     & CQ2 & 10.30 & 9.49 & 11.13 & 10.78 & 10.40 & 1.64 & 108 &\\
      \cline{3-9}
     & C3,Q3& \multicolumn{7}{c|}{not bonded} &\\
      \cline{3-9}
     & CQ4 & 12.77 & 10.54 & 12.19 & 11.02 & 11.60 & 2.23 & 186 &\\ \hline 
    
    FC4b & C1,Q1 & \multicolumn{7}{c|}{not bonded} & 
    \multirow{4}{*}{\parbox{2cm}{SEM}}  \\
    \cline{3-9}
     & CQ2 & 5.1 & 5.6 & 7.3 & 7.1 & 6.3 & 2.2 & 183 & \\
      \cline{3-9}
     & C3,Q3& \multicolumn{7}{c|}{not bonded} & \\
      \cline{3-9}
     & CQ4 & 7.5 & 6.3 & 7.4 & 6.3 & 6.9 & 1.2 & 100 & \\ \hline 
    
    FC5 & C1,Q1 & \multicolumn{7}{c|}{not bonded} & 
    \multirow{4}{*}{\parbox{2cm}{SEM}}  \\
    \cline{3-9}
     & CQ2 & 8.9 & 5.6 & 7.7 & 6.3 & 7.1 & 3.3 & 275 & \\
     & CQ3 & 8.3 & 7.4 & 6.5 & 6.3 & 7.1 & 2.0 & 150 &\\
     \cline{3-9}
     & C4,Q4 & \multicolumn{7}{c|}{not bonded} &\\ \hline 
    
    FC6 & CQ1 & 7.9 & 6.4 & 7.6 & 7.0 & 7.2 & 1.5 & 125 & 
    \multirow{4}{*}{\parbox{2cm}{SEM}}  \\
     & CQ2 & 8.7 & 8.0 & 8.7 & 6.3 & 7.3 & 2.4 & 200 & \\
     & CQ3 & 9.8 & 8.0 & 9.8 & 7.1 & 8.0 & 2.7 & 225 & \\
     & CQ4 & 8.2 & 7.0 & 8.2 & 7.7 & 8.4 & 1.2 & 100 & \\ \hline 
    \end{tabular}
    \caption{Data of the distance at each of the four corners of each flip-chip module. SEM: scanning electron microscope. \label{tableSI_chip}}
\end{table}

Note that the different flip-chip modules listed in Table \ref{tableSI_chip} were meant to investigate various device designs and flip-chip fabrication steps. Starting from the investigation of basic devices like resonators, we gradually built more complex circuitry as we integrated new learning from each experiment. Modules from FC4a onwards focus primarily on the investigation of single- and multi-qubit devices. The coherence and fidelity results shown in this work are from FC4b and FC5.

\subsection{Transition Temperatures of the Flip-Chip Connections \label{SI:transitiontemperatures}}
We characterise the resistance of a daisy chain of bonded bump pairs constituting the flip-chip material stack employed in this work. The resistance is estimated via conventional lock-in amplified 4-wire voltage sensing under current bias ($f=18$\,Hz). The cryogenic setup for this measurement is based on a cryofree Helium-3 sorption fridge (Oxford Instruments Heliox). The temperature of the sample space is slowly swept at a rate of $0.1\,$K/min and measured by a pair of RuO$_2$ and Cernox thermometers, both recently (2021) calibrated against a primary thermometer (Al-based Coulomb blockade) in the $1.6\,$K - $30\,$K temperature range.

\begin{figure}
    \centering
    \includegraphics{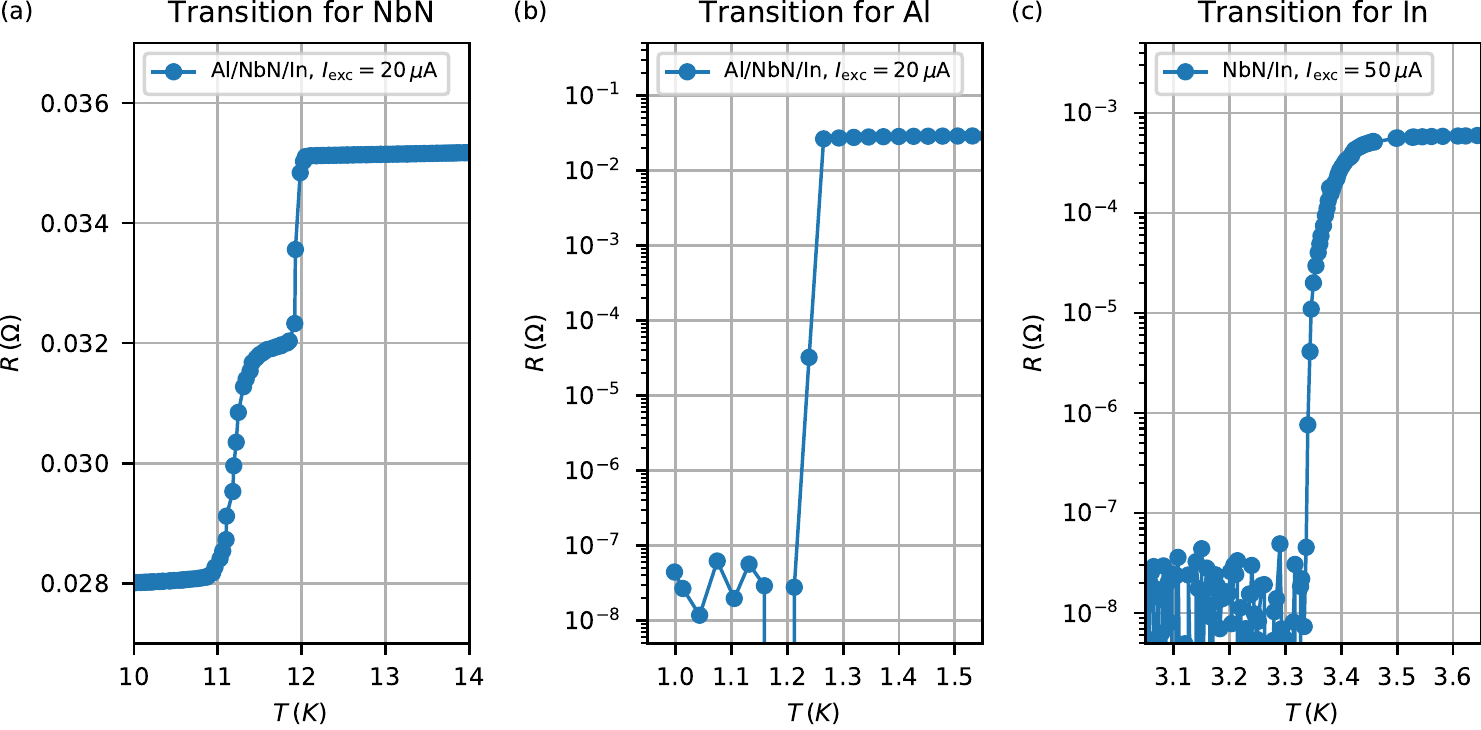}
    \caption{\textbf{Resistance measurements of the flip-chip connection}. The temperature range on the horizontal axes focus on the superconducting transition regions of NbN (a), Al (b), and In (c). The displayed resistance value corresponds to the resistance per bump or daisy chain unit. The vertical scales of (a) is linear, (b) and (c) are logarithmic. The data for (a) and (b) are obtained from measurements of the same Al/NbN/In material stack, while (c) is obtained from a separate NbN/In material stack.\label{figSI_Tc}}
\end{figure}

Figure \ref{figSI_Tc} shows the resistance per bump as a function of temperature on three temperature ranges corresponding to the superconducting transition regions of Al, In, and NbN. The data were obtained from the measurement of two different material stacks: Al/NbN/In and NbN/In.

For the Al/NbN/In stack, the trace was measured from a daisy-chain structure containing $N = 1200$ bumps under current excitation $I_\mathrm{exc}$ of 20\,$\upmu$A rms. Superconducting transitions $T_\mathrm{c}$ of the NbN structures are well visible in the $11-12\,$K temperature range (see fig.\,\ref{figSI_Tc}(a)), whereas analogous In bump features are not readily visible (not shown in fig.\,\ref{figSI_Tc}) due to their negligible resistance compared to the normal-state resistance of the Al strip (as high as $30\,$m$\Omega$ at $1.3\,$K). The latter turns superconducting below $\sim1.2\,$K (fig.\,\ref{figSI_Tc}(b)), with measured resistance reaching nonzero level of $45\,$n$\Omega$ per bump corresponding to common-mode pickup of the voltage preamplifier (NF corporation LI-75A: -130\,dB CMRR at the frequency of interest against $175\,\Omega$ resistance to measurement common in our setup). The measurement uncertainty below $1.2\,$K provides an upper bound to the resistance of the Al/NbN/In stack to be approximately $50\,$n$\Omega$ per bump or daisy chain unit.

For the NbN/In stack, the trace was measured from a daisy-chain structure containing $N=1920$ bumps under $I_\mathrm{exc}$ of $50\,\upmu$A rms, resulting in measurement uncertainty of approximately $20\,$n$\Omega$ per bump (below $\sim3.3\,$K). In this case, $T_\mathrm{c}$ of the In bumps is well visible at $\sim3.35\,$K (with normal-state resistance $0.6\,$m$\Omega$ per unit) due to the absence of any lower $T_\mathrm{c}$ superconducting components (see fig.\,\ref{figSI_Tc}(c)). Notably, an improved bonding setup with $5\,\Omega$ resistance to measurement common allows for a CMRR limit $<$n$\Omega$ per bump, well below the measurement uncertainty.

%%%%% FIGURE 
%%%%%%

\section{Two-Qubit Flip-Chip Device \label{SI:2qcoupler}}

%%%%% TABLE
\begin{table}[!h]
    \footnotesize
    \centering
    \begin{tabular}{ |c|>{\centering}p{3cm}|>{\centering}p{3cm}|c| }
        \hline
         \textbf{Parameter} & \textbf{q1} & \textbf{q2} & \textbf{coupler}\\ \hline
         \multirow{2}{*}{$f_{01}\,$(MHz)} & 
         \multirow{2}{*}{4146} & 
         \multirow{2}{*}{4776} & 7920 ($\Phi_\mathrm{c}=0$) \\
         & & & 5554 ($\Phi_\mathrm{c}=0.34\,\Phi_0$) \\ \hline
         $\alpha\,$(MHz) & -217 & -210 & -75 \\ \hline
         $E_\mathrm{J}/E_\mathrm{C}$ & 63 & 84 & \\ \hline
         $E_\mathrm{C}/h\,$(MHz) & 194 & 192 & \\ \hline
         $f_\mathrm{r}\,$(MHz) & 6210 & 6370 & \\ \hline
         $\kappa_\mathrm{r}/2\pi\,$(MHz) & 0.63 & 0.76 & \\ \hline
         $g_\mathrm{q,c}\,$(MHz) & 27 & 30 & \\ \hline 
         $g_\mathrm{q,r}\,$(MHz) & 95 & 115 & \\ \hline 
         $T_1\,(\upmu$s) ($\Phi_\mathrm{c}=0$) & $(79 \pm 11)$ & $(39\pm6)$ & \\ 
         $T_1\,(\upmu$s)  ($\Phi_\mathrm{c}=0.34\,\Phi_0$) & $(55 \pm 11)$  & $(36\pm6)$ & \\ \hline
         $T_2^*\,(\upmu$s) ($\Phi_\mathrm{c}=0$) & $(90 \pm 25)$ & $(67\pm12)$ & \\ 
         $T_2^*\,(\upmu$s)  ($\Phi_\mathrm{c}=0.34\,\Phi_0$) & $(62 \pm 20)$  & $(59\pm14)$ & \\ \hline   
         $T_\mathrm{p}\,(\upmu$s)(XY) & 248 & 187 & \\ \hline
         $T_\mathrm{p}\,(\upmu$s)(read) & 119 & 40 & \\ \hline
         $T_\mathrm{p}\,(\upmu$s)(read+XY) & 80 & 33 & \\ \hline
         $d\,(\upmu$m)/$\Delta d\,(\upmu$m) & \multicolumn{3}{c|}{6.9/1.2} \\ \hline
           
    \end{tabular}
    \caption{\textbf{Device parameters for the two-qubit flip-chip device} shown in fig.\,2(a) of the main text. Except for the coupler's anharmonicity (obtained from simulation), all values are either measured or inferred from measurement data. $f_{01}$, $\alpha$, and $f_\mathrm{r}$ are the frequencies corresponding to the qubit $|0\rangle$-to-$|1\rangle$ transition, anharmonicity, and the readout resonator frequency. $E_\mathrm{J}$ and $E_\mathrm{C}$ are the Josephson and charging energies inferred from the measured $f_{01}$ and $\alpha$. $\kappa_\mathrm{r}/2\pi$ is the decay rate of the resonator to the transmission line estimated from the full-width at half-maximum of the readout resonator absorption tone. $g_\mathrm{q,c}$ is the qubit--coupler coupling strength (see fig.\,2[d] of the main text), and $g_\mathrm{q,r}$ is the qubit--resonator coupling strength inferred from the shift of the resonator frequency from low to high measurement power (`punch-out' measurement, see Ref.\,\cite{S_Blais2021}). $T_\mathrm{p}$(XY/read) is the estimated Purcell-decay limited lifetime of the qubit due to relaxation via the XY-line/readout resonator, see text for more details. Values for $T_1$ and $T_2^*$ are the average and standard deviation of the statistics obtained over a certain period of time (17 hours for the data taken at $\Phi_\mathrm{c}=0$, and 36 hours for that at $\Phi_\mathrm{c}=0.34\,\Phi_0$). $d$ and $\Delta d$ are, respectively, the average interchip spacing and chip tilt of the flip-chip module. \label{tableSI_1}}
\end{table}

\subsection{Device Parameters}

Table \ref{tableSI_1} summarises parameters of the two-qubit flip-chip device shown in fig.\,2(a) of the main text. Of particular interest is the estimated Purcell-decay limit to the qubit lifetime due to relaxation via the readout resonator and the XY-line, i.e., $T_\mathrm{p}$(read+XY), see Section \ref{SI:purcell} for more detail. The limit due to the readout resonator, $T_\mathrm{p}$(read), is calculated from measured parameters. The value of the XY coupling capacitance, required to calculate the corresponding Purcell-decay limit, $T_\mathrm{p}$(XY), is obtained from a simulation using the measured average interchip spacing of $6.9\,\upmu$m. While this can only provide an estimate, the calculated Purcell limit due to both readout and XY elements is close to the maximum measured $T_1$ values of both qubits. 

The measured values of $T_1$ and $T_2^*$ are lower when the coupler is biased at $\Phi_\mathrm{c}=0.34\,\Phi_0$, i.e., when the coupler frequency is closer to the qubit frequencies, and where it is also more susceptible to magnetic-flux noise, compared to at $\Phi_\mathrm{c}=0$. This suggests non-negligible qubit relaxation and dephasing dynamics due to the coupler, an issue currently under investigation. 

\begin{figure}
    \centering
    \includegraphics{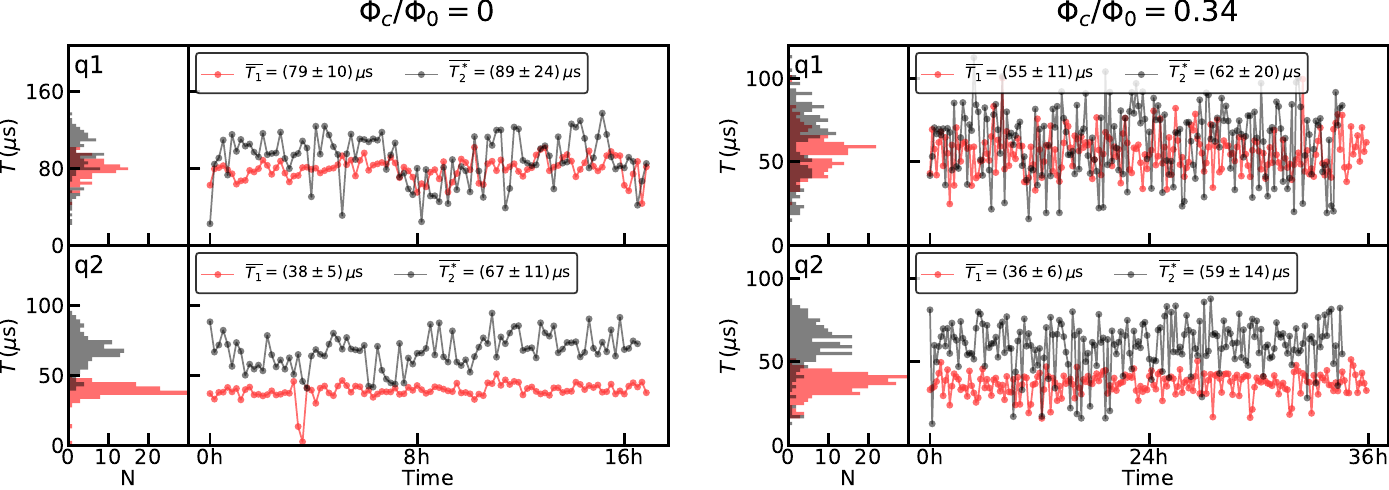}
    \caption{\textbf{Temporal fluctuations of $T_1$ and $T_2^*$} of qubits 1 and 2 in fig.\,2 of the main text\label{figSI_2qcoherence}.}
\end{figure}

Figure \ref{figSI_2qcoherence} shows the time-series and histograms of the fluctuations of both $T_1$ and $T_2^*$ values. The numbers for $T_1$ and $T_2^*$ quoted in Table \ref{tableSI_1} correspond to the average and standard deviation of these values.

%%%%% FIGURE 
%%%%%%

\subsection{Randomised Benchmarking Results}
\begin{figure}
    \centering
    \includegraphics{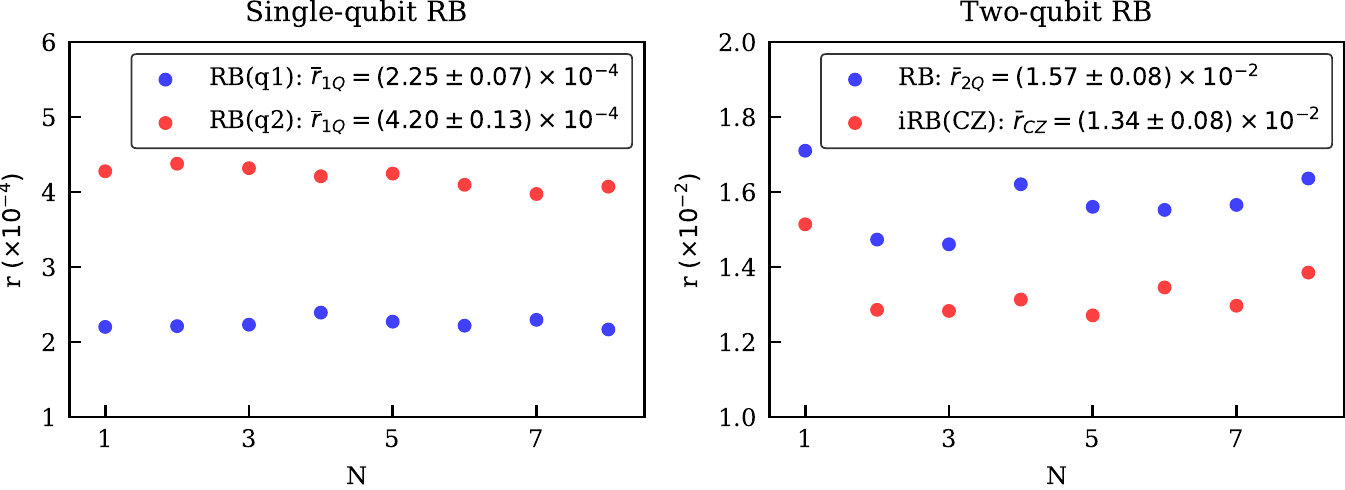}
    \caption{\textbf{Error rates $r$} for single-qubit (a) and two-qubit (b) gates obtained by randomised benchmarking with 40 random sequences repeated over 8~hours and 10~hours, respectively. 
    Data for reference RB ($\bar{r}_\mathrm{1Q}$, $\bar{r}_\mathrm{2Q}$) are average physical gate errors. Numbers quoted in the legend are mean and standard deviation values.\label{figSI_repeatedRB}}
\end{figure}

In addition to the reference single-qubit RB results shown in fig.\,3(a) of the main text, 
Table \ref{tableSI_2} lists the individual single-qubit gate errors extracted from interleaved RB experiments. Figure \ref{figSI_repeatedRB} shows the fluctuations of the error rates extracted from the RB experiments over 8~hours (single-qubit RB) and 10~hours (two-qubit RB) without any further gate calibrations in between. 
%%%%% TABLE
\begin{table}
    \footnotesize
    \centering
    \begin{tabular}{ |c|>{\centering}p{0.9cm}|>{\centering}p{0.9cm}|>{\centering}p{0.9cm}|>{\centering}p{0.9cm}|>{\centering}p{0.9cm}|>{\centering}p{0.9cm}|>{\centering}p{0.9cm}|>{\centering}p{0.9cm}|>{\centering}p{0.9cm}|c |} 
    \hline
        & Ref. & $I$ & $X_{\pi}$   &-$X_{\pi}$ &$X_{\pi/2}$ &-$X_{\pi/2}$ &  $Y_{\pi}$ &  -$Y_{\pi}$ & $Y_{\pi/2}$ &  -$Y_{\pi/2}$  \\ \hline
        $\bar{r}_\mathrm{q1}$($\times 10^{-4}$) & 1.8 & 1.6 & 2.3 & 2.4 & 1.9 & 2.8 & 4.5 & 3.4 & 2.9 & 2.8  \\ 
        $\bar{r}_\mathrm{q2}$($\times 10^{-4}$) & 5.1 & 3.0 & 4.6 & 5.9 & 4.6 & 6.4 & 8.6 & 5.4 & 4.5 & 5.3 \\
        \hline
    \end{tabular}
    \caption{\textbf{Average error rates of single-qubit gates} of qubits in fig.\,2(a) of the main text estimated via interleaved RB.\label{tableSI_2}}
\end{table}
%%%%%

%%%%% FIGURE 
%%%%%%

\subsection{Change in Device Parameters for Off-Target $d$\label{SI:2qparam_chipspacing}}
\begin{figure}
    \centering
    \includegraphics{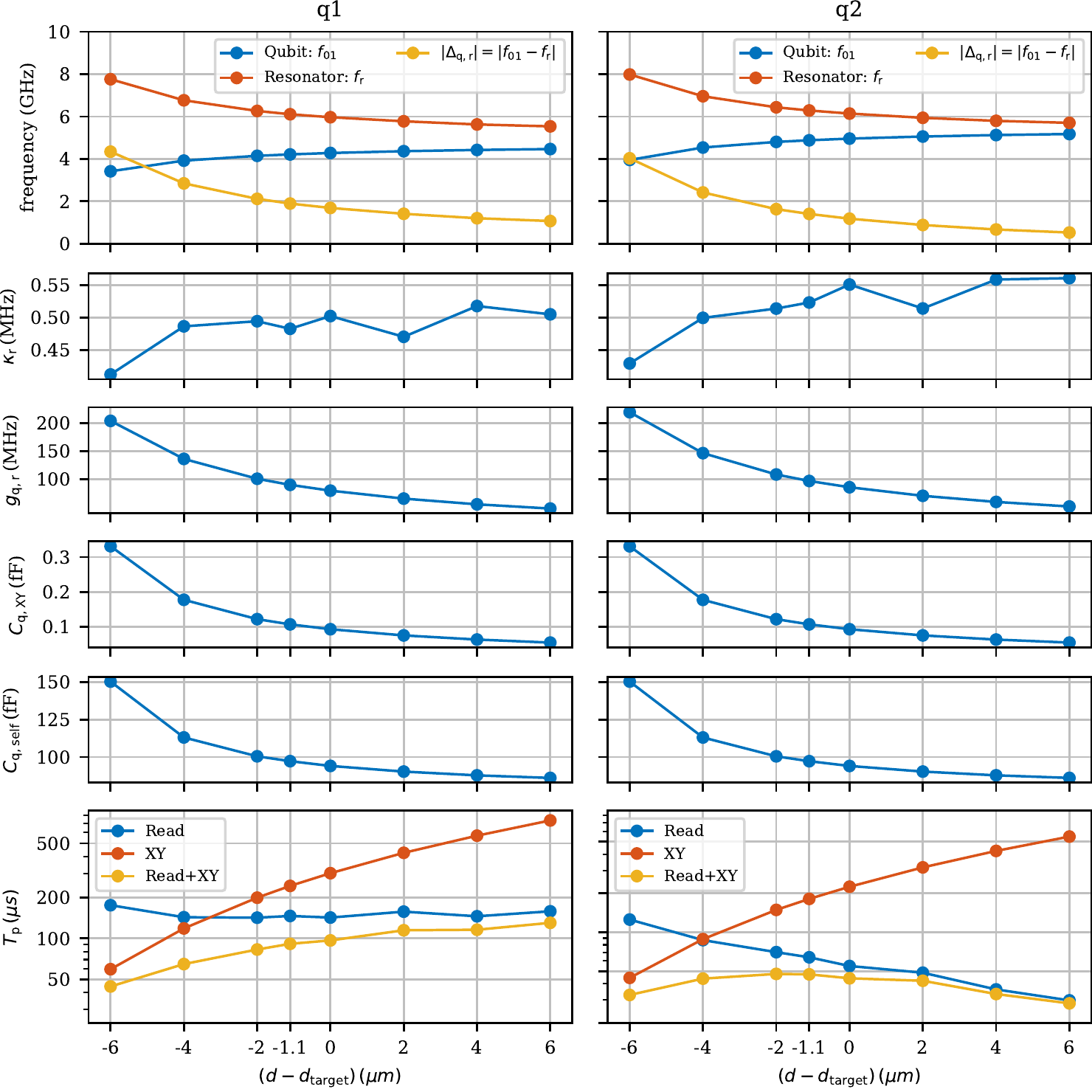}
    \caption{\textbf{Effect of off-target interchip spacing} on the device parameters of the two-qubit flip-chip device shown in fig.\,2(a) of the main text. Plots in the same row (column) share the same y-axis (x-axis). Here $d_\mathrm{target}=8\,\upmu$m is the target value for interchip spacing used during the design process. The mark for $d-d_\mathrm{target}=-1.1\,\upmu$m corresponds to the actual value of interchip spacing achieved for this device (i.e., $d=6.9\,\upmu$m). \label{figSI_purcellchipspacing}}
\end{figure}

Figure \ref{figSI_purcellchipspacing} shows the change in simulated parameters of the two-qubit flip-chip device for larger deviation of interchip spacing from the target design value ($d_\mathrm{target}=8\,\upmu$m). The purpose is to give the reader an idea of how the various parameters behave under such circumstances.

Apart from the interchip spacing value $d$, the rest of the device geometries are fixed. The value of $E_\mathrm{J}$ for each qubit is determined from the measured qubit frequency and anharmonicity values for this device. We carried out the electromagnetic simulation using the ANSYS software suite, and calculated the device parameters using the techniques outlined in Section \ref{SI:devsim} of this Supplementary Material.

First, we note that the qubit-XY coupling capacitance $C_\mathrm{q,XY}$ increases with decreasing $d$, similar to the increase in the capacitance of a parallel-plate capacitor when the plates are nearer to each other. The reasoning for the qubit self-capacitance $C_\mathrm{q, self}$ is more subtle, as there are contributions from both Q-chip (ground plane) and C-chip (ground plane, resonator, XY). In our case, there is a substantial increase in the C-chip contribution as both chips are brought closer together, which leads to the behaviour shown in fig.\,\ref{figSI_purcellchipspacing}.

The opposing trends between the qubit and resonator frequencies are more interesting. For the qubit, its inductive energy is dominated by the junction Josephson energy $E_\mathrm{J}$, which, in our case, is a fixed value. As the chips are brought closer, the increase in $C_\mathrm{q,self}$ leads to lower charging energy $E_\mathrm{C}$, and consequently lower $f_{01}$. Similarly, the resonator capacitance $C_\mathrm{r}$ increases with decreasing $d$. Crucially though, the inductive energy of the resonator, which is a transmission line device, is stored in its geometric inductance $L_\mathrm{r}$. In a flip-chip environment, $L_\mathrm{r}$ becomes \textit{smaller} with decreasing $d$, and the overall effect is such that the resonator frequency ($f_\mathrm{r}=1/(2\pi\sqrt{L_\mathrm{r}C_\mathrm{r}})$) increases with decreasing $d$.

The behaviour of the Purcell limits [$T_\mathrm{p}(\mathrm{XY})$, $T_\mathrm{p}(\mathrm{read})$] as a function of $d$ are mainly determined by the interplay between the various parameters described in eqs.\,(\ref{eq:purcelldrive}), and (\ref{eq:purcellreadout}). As a result, they could behave differently for different parameter combinations. This difference is clearly visible in comparing the response of the combined Purcell limit $T_\mathrm{p}(\mathrm{read+XY})$ between qubit 1 and qubit 2 as shown in fig.\,\ref{figSI_purcellchipspacing}.

Finally, we note that even if we achieved an on-target interchip spacing, qubit 2 is Purcell-limited to $\approx\!\!37\,\upmu$s. As discussed previously, this is due to a smaller-than-expected qubit--resonator detuning $|\Delta_\mathrm{q,r}|$ caused by an off-target junction size in this fabrication run.

%%%%% FIGURE 
%%%%%%

\section{Single-Qubit Flip-Chip Device\label{SI:sq}}

\subsection{Device Parameters}
\begin{figure}
    \centering
    \includegraphics{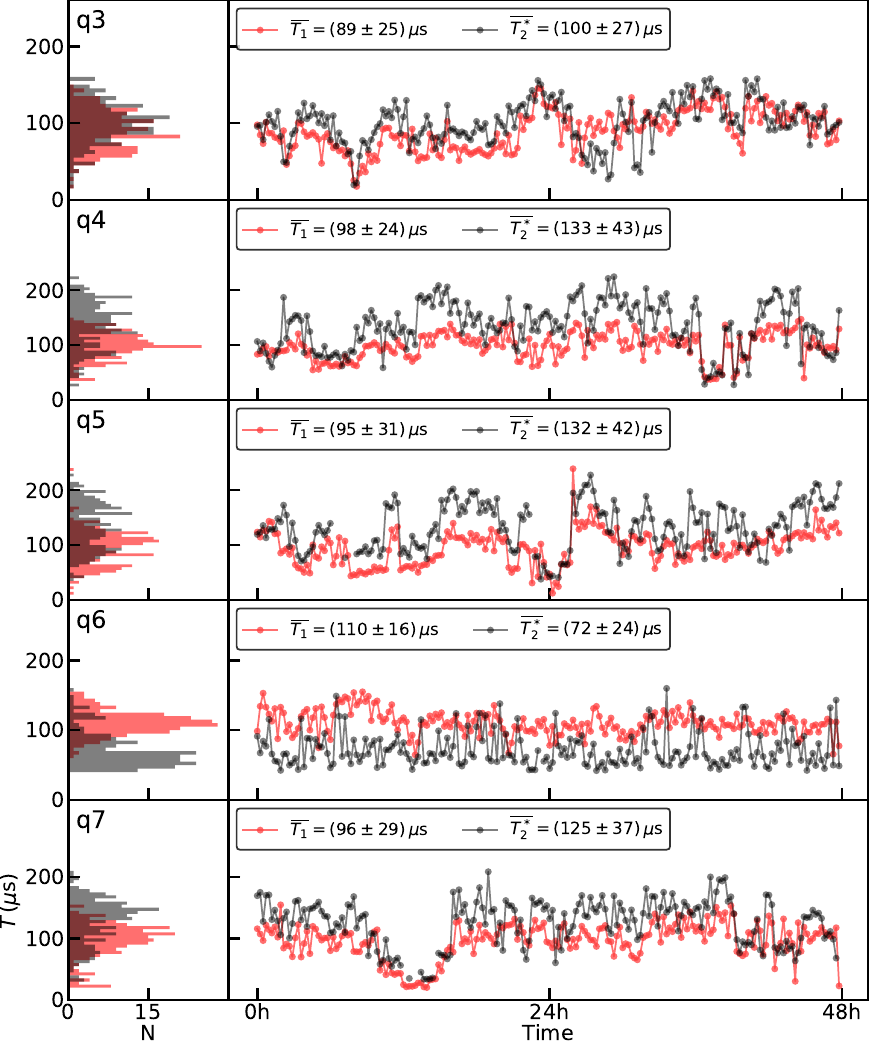}
    \caption{\textbf{Temporal fluctuation of $T_1$ and $T_2^*$} from the single-qubit devices in Table\,\ref{tableSI_3}.}
    \label{figSI_sqtimeseries}
\end{figure}

Table \ref{tableSI_3} summarises the parameters of the single-qubit flip-chip devices discussed in Section 4.3 of the main text. Here, the measured $T_1$ values are similar to what we regularly achieve in single-chip devices (and $T_1$ is not limited by Purcell decay). 
Figure \ref{figSI_sqtimeseries} shows the histogram and the time-series of $T_1$ and $T_2^*$ from each single-qubit device.

%%%%% TABLE
\begin{table}
    \footnotesize
    \centering
    \begin{tabular}{ |c|>{\centering}p{2cm} |>{\centering}p{2cm}|>{\centering}p{2cm}|>{\centering}p{2cm}|c| } 
    \hline
        \textbf{Parameter} & \textbf{q3} & \textbf{q4} & \textbf{q5} & \textbf{q6} & \textbf{q7} \\ \hline
        $f_{01}\,$(MHz) & 3974 & 4454 & 4890 & 3854 & 4887  \\ \hline
        $\alpha\,$(MHz) & -172 & -178 & -203 & -222 & -214 \\ \hline
        $E_\mathrm{J}/E_\mathrm{C}$ & 87 & 100 & 93 & 54 & 85 \\ \hline
        $E_\mathrm{C}/h\,$(MHz) & 157 & 164 & 186 & 196 & 195 \\ \hline
        $f_\mathrm{r}\,$(MHz) & 5727 & 6185 & 6691 & 5887 & 6720 \\ \hline
        $\kappa_\mathrm{r}/2\pi$(MHz) & 0.23 & 0.19 & 0.49 & 0.25 & 0.37 \\ \hline
        $g_\mathrm{q,r}\,$(MHz) & 85 & 84 & 79 & 80 & 81 \\ \hline 
        $T_1\,(\upmu$s) & $(89\pm25)$ & $(98\pm24)$ & $(95\pm31)$ & $(110\pm 16)$ & $(96\pm29)$ \\  \hline
        $T_2^*\,(\upmu$s) & $(100\pm27)$ & $(133\pm43)$ & $(132\pm42)$ & $(72\pm24)$ & $(125\pm37)$ \\ \hline   
        $T_\mathrm{p}\,(\upmu$s)(read) & 294 & 356 & 169 & 411 & 220 \\ \hline
        $d\,(\upmu$m)/$\Delta d\,(\upmu$m) & \multicolumn{3}{c|}{7.1/3.3} &\multicolumn{2}{c|}{7.1/2.0}\\
        \hline
    \end{tabular}
    \caption{\textbf{Parameters for the single-qubit flip-chip devices} discussed in Section 4.3 of the main text. A description of each parameter can be found in the caption of Table \ref{tableSI_1}. \label{tableSI_3}}
\end{table}

%%%%% FIGURE 
%%%%%%

\section{Coherence-time limits for gate fidelities}
\label{SI:coherence-limits}

To find the coherence-time limits for the gate fidelities discussed in Section~5.1 of the main text, we use the results derived in Ref.\,\cite{S_Abad2021}. There, it is shown that, to first order in the ratio $\tau / T$ of the gate time $\tau$ and the time-scale $T$ of the dissipation, the average gate fidelity for an $N$-qubit gate affected by relaxation and dephasing is
\begin{equation}
\mathcal{F}_{N, \mathrm{inc}} = 1  -\frac{d \tau}{2(d+1)} \sum_{k=1}^N \left( \frac{1}{T_1^{(k)}} + \frac{1}{T_\varphi^{(k)}} \right),
\label{eq:fidelity-N-qubit-gate}
\end{equation}
where $d = 2^N$ is the dimension of the Hilbert space for the $N$ qubits, $T_1^k$ is the relaxation time of qubit $k$, and $T_\varphi^k$ is the pure dephasing time of qubit $k$. Note that this expression is the same for all $N$-qubit gates, as long as their dynamics are restricted to the computational subspace.

Using eq.\,(\ref{eq:fidelity-N-qubit-gate}) and $1 / T_2^* = 1 / 2 T_1 + 1 / T_\varphi$, we have that the single-qubit ($N = 1$) gate fidelity is given by
\begin{equation}
\mathcal{F}_\mathrm{1Q, inc} = 1 - \frac{\tau_\mathrm{1Q}}{3} \left( \frac{1}{2 T_1} + \frac{1}{T_2^*} \right)
\label{eq:F1Qinc}
\end{equation}
and the two-qubit ($N = 2$) gate fidelity is
\begin{equation}
\mathcal{F}_\mathrm{2Q, inc} = 1 - \frac{2 \tau_\mathrm{2Q}}{5} \left( \frac{1}{2 T_1^{(\mathrm{q1})}} + \frac{1}{T_2^{*(\mathrm{q1})}} + \frac{1}{2 T_1^{(\mathrm{q2})}} + \frac{1}{T_2^{*(\mathrm{q2})}} \right) .
\label{eq:F2Qinc}
\end{equation}

Since the CZ gate as implemented here has the system temporarily leaving the computational subspace when the initial state is $\ket{11}$, this expression will be slightly modified~\cite{S_Fried2019}. However, the difference in the impact of decoherence on $\ket{11}$ and $\ket{02}$ is not large, so eq.~(\ref{eq:F2Qinc}) remains a good approximation.

Inserting the gate time $\tau_\mathrm{1Q} = 20\,$ns and the average values for $T_1$ and $T_2^*$ at $\Phi_\mathrm{c}=0.34\,\Phi_0$ for qubits 1 and 2 from Table\,\ref{tableSI_1} in eq.\,(\ref{eq:F1Qinc}) yields that the coherence-time limit for single-qubit gates was $\mathcal{F}_\mathrm{q1, inc} \sim 99.983 \%$ for qubit 1 and $\mathcal{F}_\mathrm{q2, inc} \sim 99.979 \%$ for qubit 2. From this, we calculated the time-coherence limit to single-qubit gate fidelity quoted in the main text, i.e., $\mathcal{F}_\mathrm{1Q,inc}=(\mathcal{F}_\mathrm{q1,inc}+\mathcal{F}_\mathrm{q2,inc})/2\sim99.98\,\%$. Similarly, inserting the gate time $\tau_\mathrm{2Q} = 295\,$ns and these coherence times in eq.\,(\ref{eq:F2Qinc}) yields that the coherence-time limit for two-qubit gates was $\mathcal{F}_\mathrm{2Q, inc} \sim 99.34 \%$.

\section{Participation Ratio Simulation\label{SI:pratiodiscussion}}
This section focuses on the simulated participation ratio $p_i$ of the different domains of a coplanar-waveguide (CPW) cross-section in the single-chip and flip-chip situations. We show that, for the typical CPW geometry, interchip spacing $d$, and lossy dielectric parameters used in this work, there is no expected degradation in the resulting $Q$-factor. By extending the analysis to small values of $d$, the $p_i$-value of one of the lossy interfaces is increased substantially, leading to an equivalent $Q$-factor that would become even worse than the equivalent single-chip $Q$-factor.

The participation ratio $p_i$ of domain $\Omega_i$ is defined as the fraction of its electric energy $w_i$ and the total electric energy $w$ of the whole domain \cite{S_Gao2008,S_Wenner2011,S_Wang2015,S_Gambetta2017,S_Calusine2018,S_Woods2019}, i.e.,
\begin{align}
    p_i = \frac{w_i}{w} , \quad\quad w_i = \int_{\Omega_i} \epsilon_\mathrm{r}(\Omega_i) \left|\vec{E}(\vec{r})\right|^2 d\vec{r}, \quad\quad w= \sum_i w_i,
\end{align}
where $\epsilon_\mathrm{r}(\Omega_i)$ is the relative permittivity of domain $\Omega_i$. Together with the loss-tangent value $\tan \delta_i$ for each domain, they can be used to calculate the equivalent $Q$-factor,
\begin{align}
    \frac{1}{Q} = \sum_i p_i \tan \delta_i. \label{eq:pratio_Qfactor}
\end{align}

%%%%% FIGURE 
\begin{figure}
    \centering
    \includegraphics{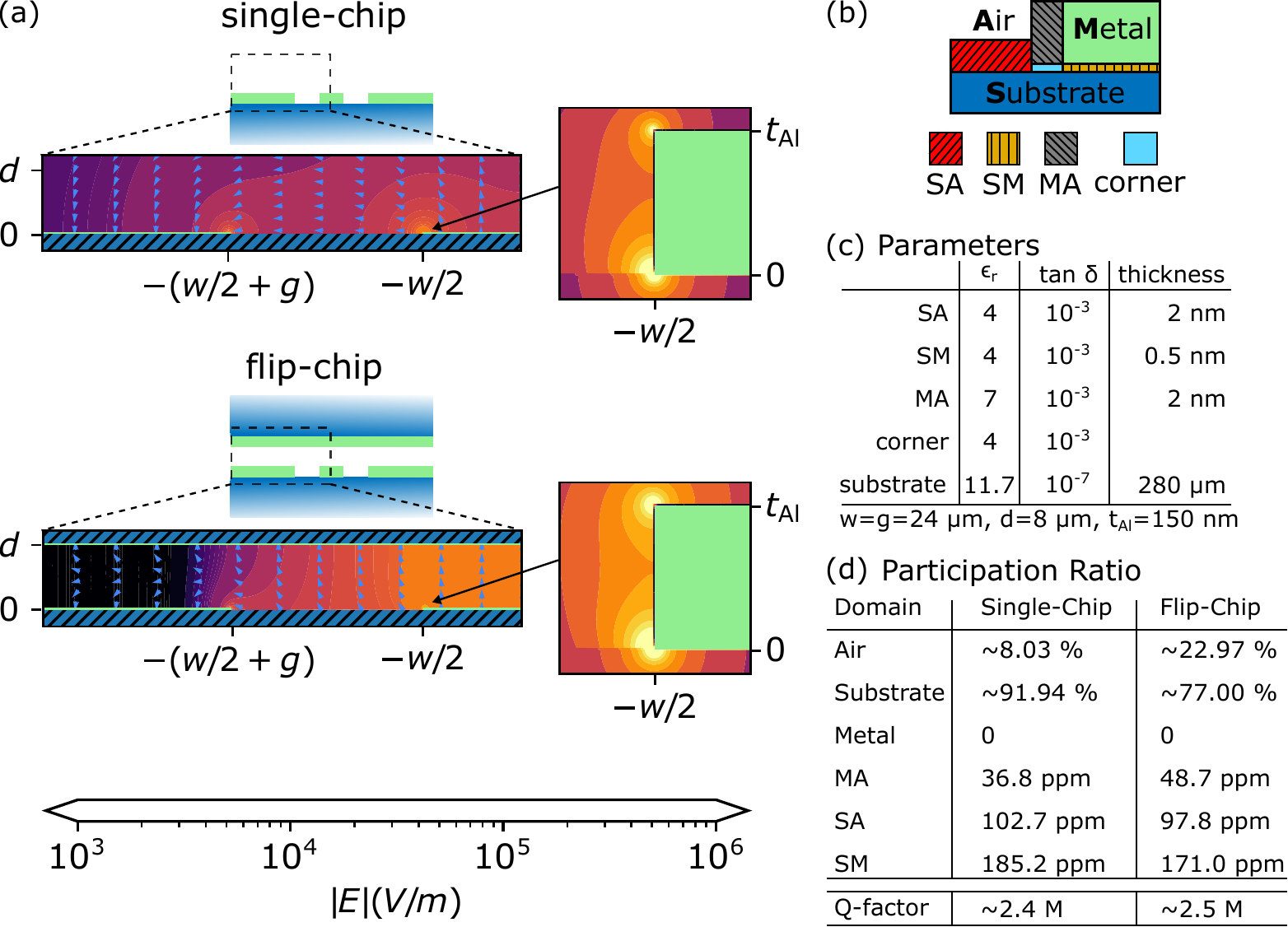}
    \caption{\textbf{Comparing simulated E-field distribution in single-chip and flip-chip geometries.} (a) Contour plots of the electrical field magnitude for the same CPW geometry in both single-chip and flip-chip environment. The color map is in logarithmic scale. The arrow indicates the direction of the E-field. (b) Designation of the various domains for the participation ratio simulation. (c) Parameters used in the simulation. (d) Calculated participation ratio of the different domains for both single-chip and flip-chip CPW geometries.\label{figSI_pratio}}
\end{figure}
%%%%%%

The E-field of the CPW cross-section is simulated using ANSYS 2D Extractor. The modelled domains are: (1) the Si chip(s) (or Substrate), (2) the Al films for the signal line and the ground plane (or Metal), (3) the lossy dielectric interfaces (Metal-Air/MA, Substrate-Air/SA, and Substrate-Metal/SM), and (4) the rest is designated as the Air, as shown in fig.\,\ref{figSI_pratio}(b). The Metal domain is set as a perfect electric conductor and assigned as either the ``Signal Line" or the ``Reference Ground". The ``Solve Options" setting for each of conductor is ``Solve on Boundary". The initial seed mesh setting is as follows: (1) a $10\,\upmu$m maximum element length for a rectangular area enclosing the CPW section (the size is $1.5a\times a$, where $a=(w+2g)$, and $w$, $g$ are respectively the width of the signal line, and the CPW signal-ground gap distance), (2) a $10\,$nm maximum element length for each $2\times2\upmu$m$^2$ rectangular area enclosing the four relevant corners of the CPW structure. The solution frequency is 5\,GHz, and the ``CG Parameter Convergence" error is 0.1\%.

The geometry and/or dielectric parameters of each domain are listed in fig.\,\ref{figSI_pratio}(c). The Air domain is considered lossless. The dielectric thicknesses and permittivity $\epsilon_\mathrm{r}$ are our current best guess based on the information we have on our processes. The latter is not an issue for the purpose of our discussion as we are interested in the change of the $p_i$ values going from the single-chip case (a bare CPW) to the flip-chip case (same CPW geometry, with an additional metal layer at a distance of $d=8\,\upmu$m). Similarly, as long as the same $\tan \delta_i$ values are used in the comparison of single-chip vs flip-chip, then it does not affect the conclusion since they are only scaling factors, see eq.\,(\ref{eq:pratio_Qfactor}).

Figure \ref{figSI_pratio}(a) shows the plots of the $E$-field magnitude in both single-chip and flip-chip situations. The additional ground plane on the top of the CPW structure redistributes the electric energy so that it appears more concentrated in the region between both chips. The table of $p_i$ in fig.\,\ref{figSI_pratio}(d) clearly shows that a larger portion of the energy is now in the Air domain. Crucially, there is redistribution in the $p_i$ of each lossy interface. Notably the MA contribution increases while both SM and SA contributions decrease. Note that the energy contribution from the domain labelled ``corner" in fig.\,\ref{figSI_pratio}(b) is divided equally between the MA and the SM domains.

Using eq.\,(\ref{eq:pratio_Qfactor}), the calculated $Q$-factor increased by $\sim\!\!4\,\%$, going from the single-chip and flip-chip case. This shows that the addition of another chip in close proximity does not necessarily lead to performance degradation, provided that the additional flip-chip fabrication steps do not add more lossy layers to the MA and the SA domains. Note that the $\sim\!\!4\,\%$ increase in $Q$-factor corresponds to $\sim\!\!4\,\upmu$s increase in equivalent $T_1$ for a 5\,GHz-qubit, which could be hard to discern in view of the much larger fluctuation in the qubit $T_1$ ($\sim\!\!20-30\,\upmu$s, see fig.\,\ref{figSI_sqtimeseries}) typically attributed to the presence of TLS impurities.

%%%%% FIGURE 
\begin{figure}
    \centering
    \includegraphics{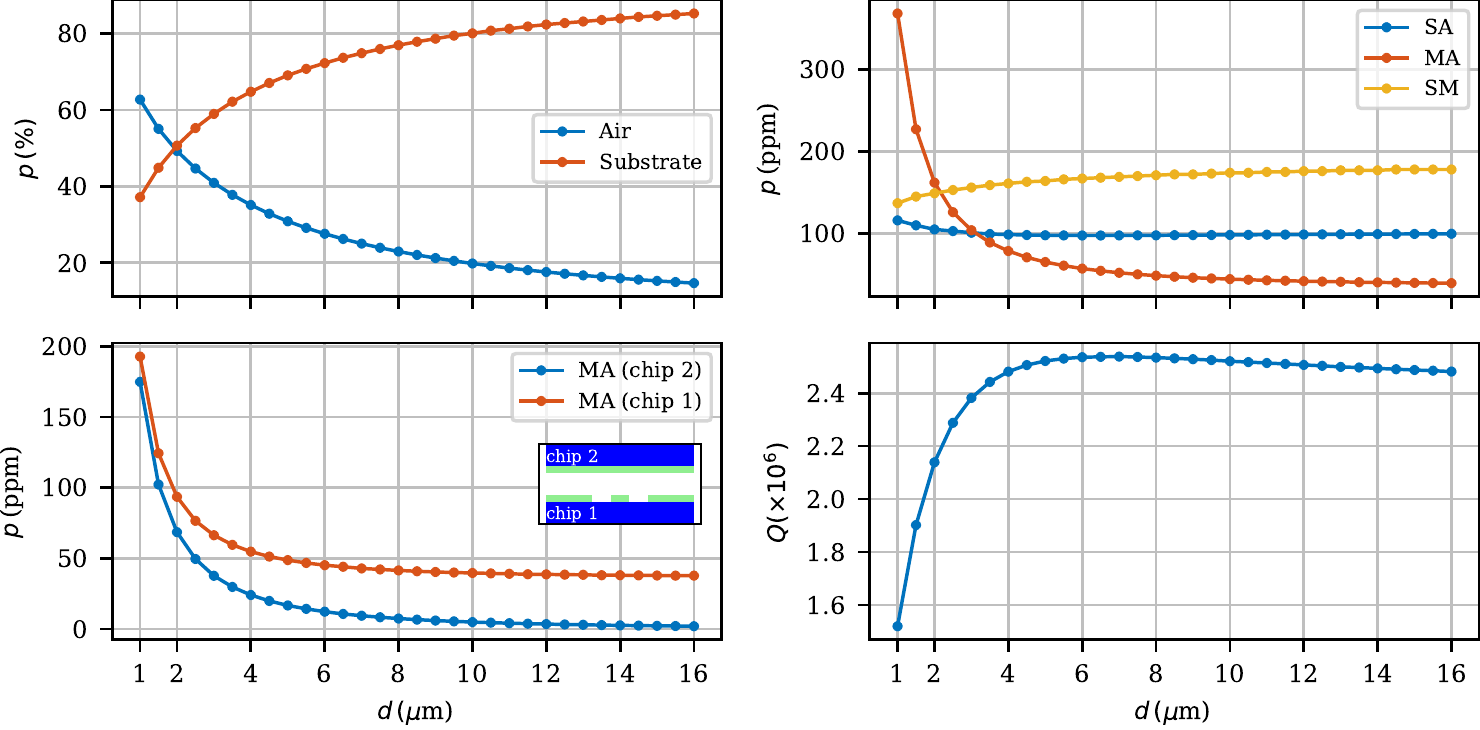}
    \caption{\textbf{Simulation of participation ratio ($p$) of the different domains and overall quality factor ($Q$)} for different interchip spacing ($d$) of a CPW geometry in a flip-chip environment. Apart from $d$, the rest of the parameters are identical to those used in fig.\,\ref{figSI_pratio}. \label{figSI_pratiochipspacing}}
\end{figure}
%%%%%%

To further illustrate the impact of the additional chip on the $p_i$ values, we ran the simulation for different values of $d$, from $1\,\upmu$m to $16\,\upmu$m. The results are shown in fig.\,\ref{figSI_pratiochipspacing}. As $d$ decreases, the Air $p_i$ increases while the Substrate $p_i$ decreases to the point that it is possible to obtain equal energy in the Air and in the Substrate (see $d=2\,\upmu$m). As this happens, the contribution from the MA region also increases substantially and eventually becomes the most dominant interface for $d$ below $\sim\!\!2\,\upmu$m. Analysing the contributions from both chips separately for large $d$ (say $d=8\,\upmu$m), the MA interface on the additional chip (labelled as chip 2 in fig.\,\ref{figSI_pratiochipspacing}) contributes about 15\,\% to the overall MA participation ratio. However, this portion increases as the chips come closer to the point that it is similar to the contribution from chip 1 at $d\sim 1\,\upmu$m. The substantial increase in contribution from the MA region for very small $d$ leads to the degradation of $Q$-factor.

A drawback of this simulation is the fact that it is not based on the complete 3D model, which would require more computational power and memory. It is possible that the 3D model would yield similar results as the transmon shape we use consist of multiple arms of CPW-like structure. Notwitstanding this issue, the $Q$-factor for the flip-chip case is $\sim\!\!4\,\%$ higher than in the single-chip case, which is perhaps an indication of further opportunities to tune the qubit geometry for an optimised $Q$-factor. Given a pre-determined value of $d$ (which is fixed by the fabrication capability), could there be a ``sweet spot" for the combination of $w$ and $g$ that would yield an optimised $Q$-factor? Do the conclusions here hold if we remove the metal layer from the additional chip, thereby allowing the $E$-field from the device to penetrate into it? Would the 3D model simulation yield similar results as the 2D model? These are interesting problems that could be addressed in future works.

\section{Sensitivity of Qubit Frequencies due to Fluctuations in $E_\mathrm{C}$ and $E_\mathrm{J}$ \label{SI:fluct}}
We describe the manner in which we arrive at a variation of $\sim\!\!4.1\,\%$ in qubit frequencies, when taking account variations in $E_\mathrm{J}$ and $E_\mathrm{C}$ (see Section 5.2 of the main text).

For a transmon qubit, the $|0\rangle$-to-$|1\rangle$ transition frequency is by Ref.\,\cite{S_Koch2007} as
\begin{equation}
    hf_{01} = \sqrt{8E_\mathrm{J}E_\mathrm{C}} - E_\mathrm{C}. 
\end{equation}
The charging energy depends on the qubit self-capacitance, i.e., $E_\mathrm{C} = e^2/(2C)$, and the Josephson energy depends on the junction critical current ($I_\mathrm{c}$) or equivalently the junction normal-state resistance $R_\mathrm{N}$, i.e., $E_\mathrm{J} = hI_\mathrm{c}/(4\pi e)$ and $I_\mathrm{c}R_\mathrm{N} = \pi\Delta/(2e)$, where $\Delta$ is the aluminium gap parameter.

Consider a transmon qubit with $f_{01}=4$\,GHz, $E_\mathrm{C}/h=200\,$MHz, and $E_\mathrm{J}/h=11000\,$MHz. For the qubit design shown in fig.\,2(a) of the main text, a variation in the interchip spacing of $1\,\upmu$m leads to a $\sim\!\!2.6\,\%$ relative change in $E_\mathrm{C}$.

A Josephson energy of $E_\mathrm{J}/h=11000\,$MHz requires a normal-state resistance $R_\mathrm{N}\sim12500\,\Omega$, which in our case, is equivalent to a junction area of $A\sim 0.022\,\upmu$m$^2$. This translates to a variation of $\sim\!\!5.5\,\%$ in the normal-state resistance $R_\mathrm{N}$ or equivalently $\sim\!\!5.5\,\%$ in $E_\mathrm{J}$, according to fig.\,4(e) of Ref.\,\cite{S_Osman2021}.

These two sources of variations are then propagated to $f_{01}$ according to the following equation:
\begin{align}
    \begin{split}
    df_{01} &= \frac{\partial{f_{01}}}{\partial{E_\mathrm{C}}}dE_\mathrm{C} + \frac{\partial{f_{01}}}{\partial{E_\mathrm{J}}}dE_\mathrm{J} \\
    & = \left(\frac{1}{2}\sqrt{\frac{8E_\mathrm{J}}{E_\mathrm{C}}}-1\right)dE_\mathrm{C} + \left(\frac{1}{2}\sqrt{\frac{8E_\mathrm{C}}{E_\mathrm{J}}}\right)dE_\mathrm{J} \\
    \frac{df_{01}}{f_{01}} &= \left(\frac{1}{2}\sqrt{\frac{8E_\mathrm{J}}{E_\mathrm{C}}}-1\right)\frac{E_\mathrm{C}}{f_{01}}\left(\frac{dE_\mathrm{C}}{E_\mathrm{C}}\right) + \left(\frac{1}{2}\sqrt{\frac{8E_\mathrm{C}}{E_\mathrm{J}}}\right)\frac{E_\mathrm{J}}{f_{01}}\left(\frac{dE_\mathrm{J}}{E_\mathrm{J}}\right).
    \end{split}
\end{align}
This results in $df_{01}/f_{01}\sim 4.1\,\%$ for the stated $f_{01}$, $E_\mathrm{C}$, $E_\mathrm{J}$, and assuming $dE_\mathrm{C}/E_\mathrm{C}\sim 2.6\,\%$ and $dE_\mathrm{J}/E_\mathrm{J}\sim5.5\%$. We note in particular that the quoted $dE_\mathrm{J}/E_\mathrm{J}$ is related to the actual standard deviation while $dE_\mathrm{C}/E_\mathrm{C}$ is estimated from the simulation, as such the variation in frequency $df_{01}/f_{01}$ should be regarded as an order of magnitude estimation.

\section*{References}
\providecommand{\newblock}{}


\begin{thebibliography}{10}
\expandafter\ifx\csname url\endcsname\relax
  \def\url#1{{\tt #1}}\fi
\expandafter\ifx\csname urlprefix\endcsname\relax\def\urlprefix{URL }\fi
\providecommand{\eprint}[2][]{\url{#2}}

\bibitem{Arute2019}
Arute F, Arya K, Babbush R, Bacon D, Bardin J~C, Barends R, Biswas R, Boixo S,
  Brandao F~G~S~L, Buell D~A, Burkett B, Chen Y, Chen Z, Chiaro B, Collins R,
  Courtney W, Dunsworth A, Farhi E, Foxen B, Fowler A, Gidney C, Giustina M,
  Graff R, Guerin K, Habegger S, Harrigan M~P, Hartmann M~J, Ho A, Hoffmann M,
  Huang T, Humble T~S, Isakov S~V, Jeffrey E, Jiang Z, Kafri D, Kechedzhi K,
  Kelly J, Klimov P~V, Knysh S, Korotkov A, Kostritsa F, Landhuis D, Lindmark
  M, Lucero E, Lyakh D, Mandr{\`{a}} S, McClean J~R, McEwen M, Megrant A, Mi X,
  Michielsen K, Mohseni M, Mutus J, Naaman O, Neeley M, Neill C, Niu M~Y, Ostby
  E, Petukhov A, Platt J~C, Quintana C, Rieffel E~G, Roushan P, Rubin N~C, Sank
  D, Satzinger K~J, Smelyanskiy V, Sung K~J, Trevithick M~D, Vainsencher A,
  Villalonga B, White T, Yao Z~J, Yeh P, Zalcman A, Neven H and Martinis J~M
  2019 {\em Nature\/} {\bf 574} 505--510 ISSN 0028-0836
  \urlprefix\url{http://www.nature.com/articles/s41586-019-1666-5}

\bibitem{Jurcevic2021}
Jurcevic P, Javadi-Abhari A, Bishop L~S, Lauer I, Bogorin D~F, Brink M,
  Capelluto L, G{\"{u}}nl{\"{u}}k O, Itoko T, Kanazawa N, Kandala A, Keefe G~A,
  Krsulich K, Landers W, Lewandowski E~P, McClure D~T, Nannicini G, Narasgond
  A, Nayfeh H~M, Pritchett E, Rothwell M~B, Srinivasan S, Sundaresan N, Wang C,
  Wei K~X, Wood C~J, Yau J~B, Zhang E~J, Dial O~E, Chow J~M and Gambetta J~M
  2021 {\em Quantum Science and Technology\/} {\bf 6} 025020 ISSN 2058-9565
  \urlprefix\url{https://iopscience.iop.org/article/10.1088/2058-9565/abe519}

\bibitem{Gong2021}
Gong M, Wang S, Zha C, Chen M~C, Huang H~L, Wu Y, Zhu Q, Zhao Y, Li S, Guo S,
  Qian H, Ye Y, Chen F, Ying C, Yu J, Fan D, Wu D, Su H, Deng H, Rong H, Zhang
  K, Cao S, Lin J, Xu Y, Sun L, Guo C, Li N, Liang F, Bastidas V~M, Nemoto K,
  Munro W~J, Huo Y~H, Lu C~Y, Peng C~Z, Zhu X and Pan J~W 2021 {\em Science\/}
  {\bf 372} 948--952 ISSN 0036-8075 (\textit{Preprint} \eprint{2102.02573})
  \urlprefix\url{https://www.sciencemag.org/lookup/doi/10.1126/science.abg7812}

\bibitem{Dunsworth2018}
Dunsworth A, Barends R, Chen Y, Chen Z, Chiaro B, Fowler A, Foxen B, Jeffrey E,
  Kelly J, Klimov P~V, Lucero E, Mutus J~Y, Neeley M, Neill C, Quintana C,
  Roushan P, Sank D, Vainsencher A, Wenner J, White T~C, Neven H, Martinis J~M
  and Megrant A 2018 {\em Applied Physics Letters\/} {\bf 112} 063502 ISSN
  0003-6951 \urlprefix\url{http://aip.scitation.org/doi/10.1063/1.5014033}

\bibitem{Andersen2021}
Andersen C~K, Remm A, Lazar S, Krinner S, Lacroix N, Christoph H, {Di Paolo} A,
  Swiadek F, Norris G~J, Hermann J, Gabureac M, Blais A, Eichler C and Wallraff
  A 2021 {\em Bulletin of the American Physical Society\/}
  \urlprefix\url{http://meetings.aps.org/Meeting/MAR21/Session/C34.1}

\bibitem{Marques2021}
Marques J~F, Varbanov B~M, Moreira M~S, Ali H, Muthusubramanian N, Zachariadis
  C, Battistel F, Beekman M, Haider N, Vlothuizen W, Bruno A, Terhal B~M and
  DiCarlo L 2022 {\em Nature Physics\/} {\bf 18} 80--86 ISSN 1745-2473
  (\textit{Preprint} \eprint{2102.13071})
  \urlprefix\url{https://www.nature.com/articles/s41567-021-01423-9}

\bibitem{Rosenberg2020}
Rosenberg D, Weber S~J, Conway D, Yost D~r~W, Mallek J, Calusine G, Das R, Kim
  D, Schwartz M~E, Woods W, Yoder J~L and Oliver W~D 2020 {\em IEEE Microwave
  Magazine\/} {\bf 21} 72--85 ISSN 1527-3342
  \urlprefix\url{https://ieeexplore.ieee.org/document/9134849/}

\bibitem{Rosenberg2017}
Rosenberg D, Kim D, Das R, Yost D, Gustavsson S, Hover D, Krantz P, Melville A,
  Racz L, Samach G~O, Weber S~J, Yan F, Yoder J~L, Kerman A~J and Oliver W~D
  2017 {\em npj Quantum Information\/} {\bf 3} 42 ISSN 2056-6387
  \urlprefix\url{http://www.nature.com/articles/s41534-017-0044-0}

\bibitem{Gold2021}
Gold A, Paquette J~P, Stockklauser A, Reagor M~J, Alam M~S, Bestwick A, Didier
  N, Nersisyan A, Oruc F, Razavi A, Scharmann B, Sete E~A, Sur B, Venturelli D,
  Winkleblack C~J, Wudarski F, Harburn M and Rigetti C 2021 {\em npj Quantum
  Information\/} {\bf 7} 142 ISSN 2056-6387 (\textit{Preprint}
  \eprint{2102.13293}) \urlprefix\url{https://www.nature.com/articles/s41534-021-00484-1}

\bibitem{Conner2021}
Conner C~R, Bienfait A, Chang H~S, Chou M~H, Dumur {\'{E}}, Grebel J, Peairs
  G~A, Povey R~G, Yan H, Zhong Y~P and Cleland A~N 2021 {\em Applied Physics
  Letters\/} {\bf 118} 232602 ISSN 0003-6951
  \urlprefix\url{https://aip.scitation.org/doi/10.1063/5.0050173}

\bibitem{Li2021a}
Li X, Zhang Y, Yang C, Li Z, Wang J, Su T, Chen M, Li Y, Li C, Mi Z, Liang X,
  Wang C, Yang Z, Feng Y, Linghu K, Xu H, Han J, Liu W, Zhao P, Ma T, Wang R,
  Zhang J, Song Y, Liu P, Wang Z, Yang Z, Xue G, Jin Y and Yu H 2021 {\em
  Applied Physics Letters\/} {\bf 119} 184003 ISSN 0003-6951 (\textit{Preprint}
  \eprint{2106.00341}) \urlprefix\url{https://aip.scitation.org/doi/10.1063/5.0068255}

\bibitem{Wu2021}
Wu Y, Bao W~S, Cao S, Chen F, Chen M~C, Chen X, Chung T~H, Deng H, Du Y, Fan D,
  Gong M, Guo C, Guo C, Guo S, Han L, Hong L, Huang H~L, Huo Y~H, Li L, Li N,
  Li S, Li Y, Liang F, Lin C, Lin J, Qian H, Qiao D, Rong H, Su H, Sun L, Wang
  L, Wang S, Wu D, Xu Y, Yan K, Yang W, Yang Y, Ye Y, Yin J, Ying C, Yu J, Zha
  C, Zhang C, Zhang H, Zhang K, Zhang Y, Zhao H, Zhao Y, Zhou L, Zhu Q, Lu C~Y,
  Peng C~Z, Zhu X and Pan J~W 2021 {\em Physical Review Letters\/} {\bf 127}
  180501 ISSN 0031-9007 (\textit{Preprint} \eprint{2106.14734})
  \urlprefix\url{https://link.aps.org/doi/10.1103/PhysRevLett.127.180501}

\bibitem{Bronn2018}
Bronn N~T, Adiga V~P, Olivadese S~B, Wu X, Chow J~M and Pappas D~P 2018 {\em
  Quantum Science and Technology\/} {\bf 3} 024007 ISSN 2058-9565
  \urlprefix\url{https://iopscience.iop.org/article/10.1088/2058-9565/aaa645}

\bibitem{Bejanin2016}
B{\'{e}}janin J~H, McConkey T~G, Rinehart J~R, Earnest C~T, McRae C~R~H, Shiri
  D, Bateman J~D, Rohanizadegan Y, Penava B, Breul P, Royak S, Zapatka M,
  Fowler A~G and Mariantoni M 2016 {\em Physical Review Applied\/} {\bf 6}
  044010 ISSN 2331-7019
  \urlprefix\url{https://link.aps.org/doi/10.1103/PhysRevApplied.6.044010}

\bibitem{Tabuchi2019}
Tabuchi Y, Tamate S and Nakamura Y 2019 {\em IEICE Transactions on
  Electronics\/} {\bf E102.C} 212--216 ISSN 0916-8524
  \urlprefix\url{https://doi.org/10.1587/transele.2018SDP0001}

\bibitem{Rahamim2017}
Rahamim J, Behrle T, Peterer M~J, Patterson A, Spring P~A, Tsunoda T, Manenti
  R, Tancredi G and Leek P~J 2017 {\em Applied Physics Letters\/} {\bf 110}
  222602 ISSN 0003-6951 (\textit{Preprint} \eprint{1703.05828})
  \urlprefix\url{http://aip.scitation.org/doi/10.1063/1.4984299}

\bibitem{Yost2020}
Yost D~R~W, Schwartz M~E, Mallek J, Rosenberg D, Stull C, Yoder J~L, Calusine
  G, Cook M, Das R, Day A~L, Golden E~B, Kim D~K, Melville A, Niedzielski B~M,
  Woods W, Kerman A~J and Oliver W~D 2020 {\em npj Quantum Information\/} {\bf
  6} 59 ISSN 2056-6387 (\textit{Preprint} \eprint{1912.10942})
  \urlprefix\url{http://www.nature.com/articles/s41534-020-00289-8}

\bibitem{Burnett2019}
Burnett J~J, Bengtsson A, Scigliuzzo M, Niepce D, Kudra M, Delsing P and
  Bylander J 2019 {\em npj Quantum Information\/} {\bf 5} 54 ISSN 2056-6387
  (\textit{Preprint} \eprint{1901.04417})
  \urlprefix\url{http://dx.doi.org/10.1038/s41534-019-0168-5}

\bibitem{Bengtsson2020}
Bengtsson A, Vikst{\aa}l P, Warren C, Svensson M, Gu X, Kockum A~F, Krantz P,
  Kri{\v{z}}an C, Shiri D, Svensson I~M, Tancredi G, Johansson G, Delsing P,
  Ferrini G and Bylander J 2020 {\em Physical Review Applied\/} {\bf 14} 034010
  ISSN 2331-7019 (\textit{Preprint} \eprint{1912.10495})
  \urlprefix\url{https://link.aps.org/doi/10.1103/PhysRevApplied.14.034010}

\bibitem{Osman2021}
Osman A, Simon J, Bengtsson A, Kosen S, Krantz P, {P Lozano} D, Scigliuzzo M,
  Delsing P, Bylander J and {Fadavi Roudsari} A 2021 {\em Applied Physics
  Letters\/} {\bf 118} 064002 ISSN 00036951 (\textit{Preprint}
  \eprint{2011.05230})
  \urlprefix\url{https://aip.scitation.org/doi/10.1063/5.0037093}

\bibitem{Niedzielski2019}
Niedzielski B~M, Yoder J~L, Ruth-Yost D, Oliver W~D, Kim D~K, Schwartz M~E,
  Rosenberg D, Calusine G, Das R, Melville A~J, Plant J and Racz L 2019
  {Silicon Hard-Stop Spacers for 3D Integration of Superconducting Qubits} {\em
  2019 IEEE International Electron Devices Meeting (IEDM)\/} (IEEE) pp
  31.3.1--31.3.4 ISBN 978-1-7281-4032-2 (\textit{Preprint} \eprint{1907.12882})
  \urlprefix\url{https://ieeexplore.ieee.org/document/8993515/}

\bibitem{Patterson2019}
Patterson A, Rahamim J, Tsunoda T, Spring P, Jebari S, Ratter K, Mergenthaler
  M, Tancredi G, Vlastakis B, Esposito M and Leek P 2019 {\em Physical Review
  Applied\/} {\bf 12} 064013 ISSN 2331-7019 (\textit{Preprint}
  \eprint{1905.05670})
  \urlprefix\url{https://link.aps.org/doi/10.1103/PhysRevApplied.12.064013}

\bibitem{Spring2021}
Spring P~A, Cao S, Tsunoda T, Campanaro G, Fasciati S, Wills J, Bakr M,
  Chidambaram V, Shteynas B, Carpenter L, Gow P, Gates J, Vlastakis B and Leek
  P~J 2022 {\em Science Advances\/} {\bf 8} 1--18 ISSN 2375-2548
  (\textit{Preprint} \eprint{2107.11140})
  \urlprefix\url{http://arxiv.org/abs/2107.11140
  https://www.science.org/doi/10.1126/sciadv.abl6698}

\bibitem{AnsysEM}
{ANSYS{\textregistered} Electromagnetics Suite, Release 2020 R1.}

\bibitem{Foxen2018}
Foxen B, Mutus J~Y, Lucero E, Graff R, Megrant A, Chen Y, Quintana C, Burkett
  B, Kelly J, Jeffrey E, Yang Y, Yu A, Arya K, Barends R, Chen Z, Chiaro B,
  Dunsworth A, Fowler A, Gidney C, Giustina M, Huang T, Klimov P, Neeley M,
  Neill C, Roushan P, Sank D, Vainsencher A, Wenner J, White T~C and Martinis
  J~M 2018 {\em Quantum Science and Technology\/} {\bf 3} 014005 ISSN 2058-9565
  \urlprefix\url{https://iopscience.iop.org/article/10.1088/2058-9565/aa94fc}

\bibitem{Gordon1999}
Gordon R~G, Liu X, Broomhall-Dillard R~N~R and Shi Y 1999 {\em MRS
  Proceedings\/} {\bf 564} 335 ISSN 0272-9172
  \urlprefix\url{http://link.springer.com/10.1557/PROC-564-335}

\bibitem{Krantz2019}
Krantz P, Kjaergaard M, Yan F, Orlando T~P, Gustavsson S and Oliver W~D 2019
  {\em Applied Physics Reviews\/} {\bf 6} 021318 ISSN 1931-9401
  (\textit{Preprint} \eprint{1904.06560})
  \urlprefix\url{http://aip.scitation.org/doi/10.1063/1.5089550}

\bibitem{Roth2017}
Roth M, Ganzhorn M, Moll N, Filipp S, Salis G and Schmidt S 2017 {\em Physical
  Review A\/} {\bf 96} 062323 ISSN 2469-9926
  \urlprefix\url{https://link.aps.org/doi/10.1103/PhysRevA.96.062323}

\bibitem{McKay2016}
McKay D~C, Filipp S, Mezzacapo A, Magesan E, Chow J~M and Gambetta J~M 2016
  {\em Physical Review Applied\/} {\bf 6} 1--10 ISSN 23317019
  (\textit{Preprint} \eprint{1604.03076})
  \urlprefix\url{http://dx.doi.org/10.1103/PhysRevApplied.6.064007}

\bibitem{Magesan2011}
Magesan E, Gambetta J~M and Emerson J 2011 {\em Physical Review Letters\/} {\bf
  106} 180504 ISSN 0031-9007
  \urlprefix\url{https://link.aps.org/doi/10.1103/PhysRevLett.106.180504}

\bibitem{Epstein2014}
Epstein J~M, Cross A~W, Magesan E and Gambetta J~M 2014 {\em Physical Review
  A\/} {\bf 89} 062321 ISSN 1050-2947 (\textit{Preprint} \eprint{1308.2928})
  \urlprefix\url{https://link.aps.org/doi/10.1103/PhysRevA.89.062321}

\bibitem{Magesan2012}
Magesan E, Gambetta J~M and Emerson J 2012 {\em Physical Review A\/} {\bf 85}
  042311 ISSN 1050-2947 (\textit{Preprint} \eprint{1109.6887})
  \urlprefix\url{https://link.aps.org/doi/10.1103/PhysRevA.85.042311}

\bibitem{Chen2018}
Chen Z 2018 {\em {Metrology of Quantum Control and Measurement in
  Superconducting Qubits}\/} Ph.D. thesis UC Santa Barbara
  \urlprefix\url{https://escholarship.org/uc/item/0g29b4p0}

\bibitem{Chow2010}
Chow J~M, DiCarlo L, Gambetta J~M, Motzoi F, Frunzio L, Girvin S~M and
  Schoelkopf R~J 2010 {\em Physical Review A\/} {\bf 82} 040305 ISSN 1050-2947
  \urlprefix\url{https://link.aps.org/doi/10.1103/PhysRevA.82.040305}

\bibitem{Lucero2010}
Lucero E, Kelly J, Bialczak R~C, Lenander M, Mariantoni M, Neeley M, O'Connell
  A~D, Sank D, Wang H, Weides M, Wenner J, Yamamoto T, Cleland A~N and Martinis
  J~M 2010 {\em Physical Review A\/} {\bf 82} 042339 ISSN 1050-2947
  (\textit{Preprint} \eprint{1007.1690})
  \urlprefix\url{https://link.aps.org/doi/10.1103/PhysRevA.82.042339}

\bibitem{Ganzhorn2020}
Ganzhorn M, Salis G, Egger D~J, Fuhrer A, Mergenthaler M, M{\"{u}}ller C,
  M{\"{u}}ller P, Paredes S, Pechal M, Werninghaus M and Filipp S 2020 {\em
  Physical Review Research\/} {\bf 2} 033447 ISSN 2643-1564 (\textit{Preprint}
  \eprint{2005.05696})
  \urlprefix\url{https://link.aps.org/doi/10.1103/PhysRevResearch.2.033447}

\bibitem{Barends2013}
Barends R, Kelly J, Megrant A, Sank D, Jeffrey E, Chen Y, Yin Y, Chiaro B,
  Mutus J, Neill C, O'Malley P, Roushan P, Wenner J, White T~C, Cleland A~N and
  Martinis J~M 2013 {\em Physical Review Letters\/} {\bf 111} 080502 ISSN
  0031-9007 (\textit{Preprint} \eprint{1304.2322})
  \urlprefix\url{https://link.aps.org/doi/10.1103/PhysRevLett.111.080502}

\bibitem{Ye2021}
Ye Y, Cao S, Wu Y, Chen X, Zhu Q, Li S, Chen F, Gong M, Zha C, Huang H~L, Zhao
  Y, Wang S, Guo S, Qian H, Liang F, Lin J, Xu Y, Guo C, Sun L, Li N, Deng H,
  Zhu X and Pan J~W 2021 {\em Chinese Physics Letters\/} {\bf 38} 100301 ISSN
  0256-307X (\textit{Preprint} \eprint{2109.05680})
  \urlprefix\url{https://iopscience.iop.org/article/10.1088/0256-307X/38/10/100301}

\bibitem{Abad2021}
Abad T, Fern{\'{a}}ndez-Pend{\'{a}}s J, Kockum A~F and Johansson G 2021
  (\textit{Preprint} \eprint{2110.15883})
  \urlprefix\url{http://arxiv.org/abs/2110.15883}

\bibitem{Zhang2020}
Zhang E~J, Srinivasan S, Sundaresan N, Bogorin D~F, Martin Y, Hertzberg J~B,
  Timmerwilke J, Pritchett E~J, Yau J~B, Wang C, Landers W, Lewandowski E~P,
  Narasgond A, Rosenblatt S, Keefe G~A, Lauer I, Rothwell M~B, McClure D~T,
  Dial O~E, Orcutt J~S, Brink M and Chow J~M 2022 {\em Science Advances\/} {\bf
  8} 1--9 ISSN 2375-2548 (\textit{Preprint} \eprint{2012.08475})
  \urlprefix\url{https://www.science.org/doi/10.1126/sciadv.abi6690}

\bibitem{Place2021}
Place A~P~M, Rodgers L~V~H, Mundada P, Smitham B~M, Fitzpatrick M, Leng Z,
  Premkumar A, Bryon J, Vrajitoarea A, Sussman S, Cheng G, Madhavan T, Babla
  H~K, Le X~H, Gang Y, J{\"{a}}ck B, Gyenis A, Yao N, Cava R~J, de~Leon N~P and
  Houck A~A 2021 {\em Nature Communications\/} {\bf 12} 1779 ISSN 2041-1723
  (\textit{Preprint} \eprint{2003.00024})
  \urlprefix\url{http://dx.doi.org/10.1038/s41467-021-22030-5}

\bibitem{Wang2021}
Wang C, Li X, Xu H, Li Z, Wang J, Yang Z, Mi Z, Liang X, Su T, Yang C, Wang G,
  Wang W, Li Y, Chen M, Li C, Linghu K, Han J, Zhang Y, Feng Y, Song Y, Ma T,
  Zhang J, Wang R, Zhao P, Liu W, Xue G, Jin Y and Yu H 2021
  (\textit{Preprint} \eprint{2105.09890})
  \urlprefix\url{http://arxiv.org/abs/2105.09890}
  
  \setcounter{firstbib}{\value{enumi}}

\end{thebibliography}

\begin{thebibliography}{10}
\expandafter\ifx\csname url\endcsname\relax
  \def\url#1{{\tt #1}}\fi
\expandafter\ifx\csname urlprefix\endcsname\relax\def\urlprefix{URL }\fi
\providecommand{\eprint}[2][]{\url{#2}}

%\setcounter{bibliography}{43}

\setcounter{enumi}{\value{firstbib}}

\bibitem{S_Krantz2019}
Krantz P, Kjaergaard M, Yan F, Orlando T~P, Gustavsson S and Oliver W~D 2019
  {\em Applied Physics Reviews\/} {\bf 6} 021318 ISSN 1931-9401
  (\textit{Preprint} \eprint{1904.06560})
  \urlprefix\url{http://aip.scitation.org/doi/10.1063/1.5089550}

\bibitem{S_AnsysEM}
{ANSYS{\textregistered} Electromagnetics Suite, Release 2020 R1.}

\bibitem{S_Probst2015}
Probst S, Song F~B, Bushev P~A, Ustinov A~V and Weides M 2015 {\em Review of
  Scientific Instruments\/} {\bf 86} 024706 ISSN 0034-6748 (\textit{Preprint}
  \eprint{1410.3365}) \urlprefix\url{http://aip.scitation.org/doi/10.1063/1.4907935}

\bibitem{S_Sank2014}
Sank D~T 2014 {\em {Fast, accurate state measurement in superconducting
  qubits}\/} Ph.D. thesis UC Santa Barbara

\bibitem{S_Koch2007}
Koch J, Yu T~M, Gambetta J, Houck A~A, Schuster D~I, Majer J, Blais A, Devoret
  M~H, Girvin S~M and Schoelkopf R~J 2007 {\em Physical Review A\/} {\bf 76}
  042319 ISSN 1050-2947
  \urlprefix\url{https://link.aps.org/doi/10.1103/PhysRevA.76.042319}

\bibitem{S_Blais2021}
Blais A, Grimsmo A~L, Girvin S~M and Wallraff A 2021 {\em Reviews of Modern
  Physics\/} {\bf 93} 025005 ISSN 0034-6861
  \urlprefix\url{https://link.aps.org/doi/10.1103/RevModPhys.93.025005}

\bibitem{S_Abad2021}
Abad T, Fern{\'{a}}ndez-Pend{\'{a}}s J, Kockum A~F and Johansson G 2021
  (\textit{Preprint} \eprint{2110.15883})
  \urlprefix\url{http://arxiv.org/abs/2110.15883}

\bibitem{S_Fried2019}
Fried E~S, Sivarajah P, Didier N, Sete E~A, da~Silva M~P, Johnson B~R and Ryan
  C~A 2019  (\textit{Preprint} \eprint{1908.11370})
  \urlprefix\url{http://arxiv.org/abs/1908.11370}

\bibitem{S_Gao2008}
Gao J, Daal M, Vayonakis A, Kumar S, Zmuidzinas J, Sadoulet B, Mazin B~A, Day
  P~K and Leduc H~G 2008 {\em Applied Physics Letters\/} {\bf 92} 152505 ISSN
  0003-6951, 1077-3118
  \urlprefix\url{http://aip.scitation.org/doi/10.1063/1.2906373}

\bibitem{S_Wenner2011}
Wenner J, Barends R, Bialczak R~C, Chen Y, Kelly J, Lucero E, Mariantoni M,
  Megrant A, O’Malley P~J~J, Sank D, Vainsencher A, Wang H, White T~C, Yin Y,
  Zhao J, Cleland A~N and Martinis J~M 2011 {\em Applied Physics Letters\/}
  {\bf 99} 113513 ISSN 0003-6951, 1077-3118
  \urlprefix\url{http://aip.scitation.org/doi/10.1063/1.3637047}

\bibitem{S_Wang2015}
Wang C, Axline C, Gao Y~Y, Brecht T, Chu Y, Frunzio L, Devoret M~H and
  Schoelkopf R~J 2015 {\em Applied Physics Letters\/} {\bf 107} 162601 ISSN
  0003-6951, 1077-3118
  \urlprefix\url{http://aip.scitation.org/doi/10.1063/1.4934486}

\bibitem{S_Gambetta2017}
Gambetta J~M, Murray C~E, Fung Y~K~K, McClure D~T, Dial O, Shanks W, Sleight
  J~W and Steffen M 2017 {\em IEEE Transactions on Applied Superconductivity\/}
  {\bf 27} 1--5 ISSN 1051-8223, 1558-2515
  \urlprefix\url{http://ieeexplore.ieee.org/document/7745914/}

\bibitem{S_Calusine2018}
Calusine G, Melville A, Woods W, Das R, Stull C, Bolkhovsky V, Braje D, Hover
  D, Kim D~K, Miloshi X, Rosenberg D, Sevi A, Yoder J~L, Dauler E and Oliver
  W~D 2018 {\em Applied Physics Letters\/} {\bf 112} 062601 ISSN 0003-6951,
  1077-3118 \urlprefix\url{http://aip.scitation.org/doi/10.1063/1.5006888}

\bibitem{S_Woods2019}
Woods W, Calusine G, Melville A, Sevi A, Golden E, Kim D, Rosenberg D, Yoder J
  and Oliver W 2019 {\em Physical Review Applied\/} {\bf 12} 014012 ISSN
  2331-7019
  \urlprefix\url{https://link.aps.org/doi/10.1103/PhysRevApplied.12.014012}

\bibitem{S_Osman2021}
Osman A, Simon J, Bengtsson A, Kosen S, Krantz P, {P Lozano} D, Scigliuzzo M,
  Delsing P, Bylander J and {Fadavi Roudsari} A 2021 {\em Applied Physics
  Letters\/} {\bf 118} 064002 ISSN 00036951 (\textit{Preprint}
  \eprint{2011.05230})
  \urlprefix\url{https://aip.scitation.org/doi/10.1063/5.0037093}

\end{thebibliography}
\end{document}